\documentclass{article}
\usepackage{epsf}
\usepackage{cite}

\reversemarginpar

\newcommand{\ret}{\hfill\break\noindent}
\newcommand{\nn}{\nonumber}

\newcommand{\ba}{\begin{eqnarray}}
\newcommand{\ea}{\end{eqnarray}}
\newcommand{\abs}[1]{\left|#1\right|}
\newcommand{\text}[1]{{\rm #1}}
\def\Pslash{\hbox{P\kern-.7em\raise.2ex\hbox{/}}}
\catcode`\@=11 % This allows us to modify PLAIN macros
\def\lsim{\mathrel{\mathpalette\@versim<}}
\def\gsim{\mathrel{\mathpalette\@versim>}}
\def\@versim#1#2{\vcenter{\offinterlineskip
        \ialign{$\m@th#1\hfil##\hfil$\crcr#2\crcr\sim\crcr } }}
\catcode`\@=12 % at signs are no longer letters
\def\Ef{E_{\hbox{\tiny F}}}
\def\ef{\varepsilon_{\hbox{\tiny F}}}
\def\pf{p_{\hbox{\tiny F}}}
\def\Tf{T_{\hbox{\tiny F}}}
\def\jE{j_{\hbox{\tiny E}}}
\def\qE{q_{\hbox{\tiny E}}}
\newcommand{\mubohr}{\ifmmode \mu_{\hbox{\tiny B}} \else $\mu_{\hbox{\tiny B}}$ \fi}
\def\unit#1#2{\ifmmode \,{\rm #1}^{#2} \else ${\rm #1}^{#2}$ \fi}

\newcommand{\longhminus}{{\cal H}_{-1/2}}
\newcommand{\longhplus}{{\cal H}_{1/2}}
\newcommand{\longhminusthreehalf}{{\cal H}_{-3/2}}

\begin{document}
\begin{flushright} Preprint USM-TH-85 \end{flushright}
%\begin{flushright} \today \end{flushright}
\medskip

\begin{center} \Large\bf The Magnetized Electron Gas in terms of Hurwitz Zeta Functions\end{center}
\vspace*{.15in} \normalsize
\begin{center}
Claudio O. Dib\footnote{cdib@fis.utfsm.cl} ~and
Olivier Espinosa\footnote{espinosa@fis.utfsm.cl}
\end{center}

\vspace*{.15in}
\begin{center}
Departamento de F\'{\i}sica \\ Universidad T\'{e}cnica Federico
Santa Mar\'{\i}a \\ Casilla 110-V \\ Valpara\'{\i}so, Chile
\end{center}

\begin{abstract}
%\baselineskip=0.7cm
%To get double-spacing uncomment the following line:
%\baselineskip=24pt
%
We obtain explicit expressions for
thermodynamic quantities of a relativistic degenerate free
electron gas in a magnetic field in terms of Hurwitz Zeta
functions. The formulation allows for systematic expansion in all
regimes. Three energy scales appear naturally in the degenerate
relativistic gas: the Fermi energy $\Ef$, the temperature $T$ and
an energy related to the magnetic field or Landau level spacing,
$eB/\Ef$. We study the cold and warm scenarios, $T\ll eB/\Ef$ and
$eB/\Ef \ll T$, respectively. We reproduce the oscillations of the
magnetization as a function of the field in the cold regime and
the dilution of them in the warm regime.

\end{abstract}

\vskip 2cm
\noindent
Corresponding Author:\\
Olivier Espinosa (espinosa@fis.utfsm.cl)\\
Departamento de F\'{\i}sica, Universidad T\'{e}cnica Federico Santa Mar\'{\i}a\\
Casilla 110-V, Valpara\'{\i}so, Chile.\\
Phone: +56(32)654-506\\
Fax: +56(32)797-656\\
\vfill

%\vspace{1 in}
%37 pages and 6 figures.

\newpage
%\noindent
%{\it Proposed running head}: MAGNETIZED ELECTRON GAS\\
%{\it Mailing address}:\\
%Claudio Dib\\
%Depto Fisica, UTFSM\\
%Casilla 110-V, Valpara\'\i so\\
%CHILE.\\
%phone: +56(32)654504,\ \
%fax: +56(32)797656,\ \
%e-mail: cdib@fis.utfsm.cl

%\baselineskip=0.7cm
%To get double-spacing uncomment the following line:
%\baselineskip=24pt
\parskip=7pt
\parindent=0pt
\section{Introduction}

The study of the thermodynamic properties of degenerate
relativistic electron gases in strong magnetic fields, as those
found in compact astrophysical objects, was started long ago
\cite{Canuto}. Although in that work the
problem was stated in quite general terms, that is, arbitrary
temperature and relativistic electrons described by Dirac's
equation, approximate analytical results were obtained only in the
limits of low temperatures and non-relativistic electrons.
While the non-relativistic approximation has some validity in
white dwarfs, it is inappropriate for the electron
gas existing inside neutron stars. In fact, in a typical neutron
star the electron density is of the order of $10^{-2}\,{\rm fm}^{-3}$,
which implies a Fermi kinetic energy $\Tf\equiv\Ef-m$ of the
order of 100 MeV. The temperature, on the other hand, is at most
a few MeV, which makes the ratio $T/\Tf$ a small
parameter and the gas quite degenerate. The relativistic case
was partially considered later on in Refs.~\cite{Oconnell,Schmid}
with regard to the problem of self-magnetization, and in
Ref.~\cite{blanford} in connection with the magnetic
susceptibility. In all these works only the dominant contributions
relevant for each case were kept in the analysis.

More recently, a renewed interest in the relativistic magnetized
electron gas, both at zero and finite temperature, has appeared
from the point of view of quantum field theory
\cite{chodos,blau,elmfors,zeitlin,cangemi}. In these studies most
of the emphasis is put on the formal aspects of the problem. For
instance, in Ref.~\cite{elmfors} an exact analytic expression is
obtained for the effective action (also called the grand potential
in statistical mechanics), which corrects an incomplete result
given earlier \cite{chodos} that had missed the de Haas -- van
Alphen oscillations that are present at low temperatures, exactly
as in the non-relativistic case. In Ref.~\cite{zeitlin} the
quantum field theory results of Ref.~\cite{elmfors} are recast,
for the case of low temperatures, in terms of sums over filled
Landau levels, which is actually the starting point of the quantum
statistical mechanical calculation developed in the earlier works
\cite{Canuto,Oconnell,Schmid,blanford}.

In this work we revisit the usual elementary statistical
mechanical approach, originally carried out by Landau
\cite{landau-paper1,landau-paper2} (see also
Ref.~\cite{landau-spI}) for a non-relativistic gas, to study a
highly degenerate relativistic free electron gas, in the presence
of a uniform magnetic field. Starting from a closed analytical
expression for the density of states for this system, we derive
explicit, simple analytical expressions for the various quantities
of thermodynamic interest, such as the density and magnetization.
The object of central interest will be the grand potential, from
which all relevant thermodynamic quantities can be computed.

It is well known that for a {\em relativistic} electron gas the Coulomb
interactions among electrons and between the electrons and any
background of positive charge that should exist in a neutral
system are small (order $\alpha$) corrections to the kinetic
energy, so we will neglect them. We will therefore work consistently
to zeroth order in QED
corrections. In particular, this means that we will take the
electron gyromagnetic ratio exactly equal to two, $g=2$.
Additionally, since we shall be solely concerned with the
degenerate regime, we neglect the contribution of positrons
altogether.
\\

We will show that the grand potential at $T=0$ has the integral
representation
 \ba\label{OmegaZero0} \Omega _0 (\Ef,B) =  - \frac{V}{4\pi^2 }\
(2e B)^{5/2} \int_0^{(\Ef^2-m^2)/2eB} ~{\frac{{\longhminus
(q)}}{{\sqrt {m^2 + 2eB~q} \,}}\,dq}, \ea
where $m$ and $e$ are the electron mass and the fundamental charge,
respectively (we use natural units, so that $\hbar = c = k_B = 1$).
Consequently the electron density at $T=0$ is
\ba\label{eqforEf0} n_0 (\Ef,B) = \frac{1}{2\pi^2} (2e B)^{3/2}\
\longhminus \left( \frac{\Ef^2-m^2}{2e B} \right). \ea
The function $\longhminus(q)$ appearing in both (\ref{OmegaZero0})
and (\ref{eqforEf0}), is a combination of Hurwitz zeta functions,
$\longhminus(q)\equiv\zeta
({-\textstyle{1 \over 2}},\left\{ q \right\}) - \zeta
({-\textstyle{1 \over 2}},q + 1) - \frac{1}{2}\sqrt{q}$,
and plays a pervading role in our study of the magnetized free
electron gas. Since the pressure is given by $P = -\Omega/V$,
the expressions (\ref{OmegaZero0}) and (\ref{eqforEf0}) furnish a
parametric representation of the equation of state at $T=0$,
$P_0=P_0(\Ef,B), n_0=n_0(\Ef,B)$.

The expressions given above correspond to zero temperature,
but, as we show later in the paper, for a system of
non-interacting fermions the {\em full} finite
temperature grand potential can be obtained from its zero
temperature counterpart.
For instance, in the degenerate regime finite temperature
corrections are usually obtained systematically from the ground
state quantities in the form of a Taylor series in powers of
$T/\mu$, where $\mu$ is the chemical potential. Take, for
example, $\Omega(T,\mu)$. As is well known, the
leading correction to the zero-temperature result generally
goes as $(T/\Ef)^2$. This will actually be the case, provided
the grand potential at $T=0$,  $\Omega_0
(\Ef)$, does not vary greatly when $\Ef$ is changed by an
amount of order $T$. However, as we will see, in our case
$\Omega_0(\Ef)$ has contributions that oscillate rapidly
with $\Ef$ due to the filling of discrete Landau levels.
Therefore the approximation above is not valid, except for
very low temperatures ($T\ll eB/\Ef$, as we will see later).
We will compute with our formalism the correct finite
temperature behavior of $\Omega(T,\mu)$ and reproduce the
previously known result that the oscillations are smoothed
out as the temperature increases.

The novelty of our approach is the use of the Hurwitz Zeta
function to deal with the discrete sums over Landau levels that
accommodate the electrons in the presence of a uniform magnetic
field. The traditional approach \cite{landau-spI} makes use of the
so-called Poisson summation formula which is of limited use if one
needs to expand the resulting expressions in different regimes,
such as for small fields. The Hurwitz zeta function approach
is much more powerful in this case, as it leads to closed
analytical expressions, which can eventually be evaluated
numerically with ease, and is suitable for expansion in any
desired regime. Although this is not the first appearance of the
Hurwitz zeta function in connection with the system being
studied \cite{blau,cangemi},
this seems to be the first time that its analytic properties are
fully put into use to unravel the thermodynamics of the
relativistic magnetized free electron gas.

The Hurwitz zeta function $\zeta(z,q)$ is defined
as the analytic extension to the whole complex $z$ plane of the
series
\ba \zeta(z,q) & = &\sum_{n=0}^{\infty} \frac{1}{(n+q)^{z}} \ea
valid for $\text{Re } z > 1$ and $q \neq 0, -1, -2,  \cdots$. The
resulting function is analytic everywhere except at $z=1$, where
it has a simple pole with unit residue (see Appendix A for
details).

In section 2 we formulate the density of states in its most
general form in terms of Hurwitz zeta functions. In section 3 we
classify the physical regimes we want to study. The core of the
thermodynamics is presented in Section 4 and particular studies of
the magnetization are shown in Section 5. Details of the
calculations and mathematical formulas are given in the
appendices.

%here figure 1 below**************************

%insert the figure in the space above

\section{The Density of States in terms of Hurwitz Functions}
\label{density-of-states}

The stationary states of a Dirac electron that moves in a uniform
magnetic field ${\bf B}$ (which we take to point along the
$z$-direction), are specified in terms of four quantum numbers. In
the gauge where  ${\bf A}({\bf x})= x\ B\ {\bf\hat y}$, these
quantum numbers are $p_z$ (momentum component along ${\bf B}$),
$p_y$ (momentum component along ${\bf A}$), a non-negative integer
$n = 0,1,2,\ldots$ that specifies the Landau level, and an integer
$\lambda=\pm 1$ which denotes the spin parallel or antiparallel to
${\bf B}$. The energy levels are independent of the quantum number
$p_y$ and are given by\footnote{For $g=2$, i.e., neglecting QED
corrections to the electron's magnetic moment.}
\cite{lippman}
\ba
  E(p_z ,j) & = & \sqrt {p_z^2  + m^2  + \left( {2e B }\right)j},
\label{rel-spectrum} \ea
where $j$ is a nonnegative integer defined by $j
=n+(\lambda+1)/2$. The energy levels (\ref{rel-spectrum}) are
highly degenerate, due to their independence on $p_y$ and their
dependence on $n$ and $\lambda$ only through the single
combination $j$. The number of states $g(p_{z},j) dp_{z}$ for
given $j$ and momentum along $\bf B$ between $p_z$ and $p_z+dp_z$,
in a system of electrons confined to a finite cubic box of volume
$V$, is given by
\ba\label{deg1} g(p_z ,j) dp_{z} = g_j \frac{{eB}}{{4\pi^2 }}V dp_{z},
\ea
with $g_j=1$ for $j=0$ and $g_j =2$ for $j \ne 0$.
Physically, this degeneracy embodies the fact that
the levels with $j=0$ accommodate only electrons with spin pointing
down, whereas all the others can have electrons spinning in either
direction. As an immediate consequence, the levels with $j=0$
alone are responsible for the paramagnetic contribution to the
magnetization. See \cite{landau-qm} for details. The density of
states in energy space is thus given by
\begin{equation}
g(E)\,dE = 2\sum\limits_{j = 0}^{\jE} {g(|p_z| ,j)\,dp_z },
\end{equation}
where the factor of 2 in front of the sum takes into account both
possible signs of $p_z$, and the integer $\jE$ corresponds to the
highest Landau level $j$ that starts at an energy less than $E$.
 From Eq.~(\ref{rel-spectrum}) it is clear that $\jE$ is the
 integer part of the quantity
\begin{equation}\label{def-qE}
\qE  = \frac{{E^2  - m^2 }}{{2e B}},
\end{equation}
denoted usually as  $\jE =\left\lfloor \qE \right\rfloor$. The
degeneracy of the levels \cite{lippman} can be expressed explicitly
by
\ba\label{gE0} g(E) &=& \frac{{V}}{{4\pi^2 }}2eB~E\sum\limits_{j =
0}^{\jE} {g_j \frac{1}{{\sqrt {E^2  - m^2  - 2e B\cdot j} }}}\\
\nonumber &=& \frac{{V}}{{4\pi^2 }}(2e B)^{1/2} E\sum\limits_{j =
0}^{\left\lfloor {\qE } \right\rfloor } {g_j \frac{1}{{\sqrt {\qE
- j} }}}.
\ea
Considering now the degeneracy $g_j$ in Eq.~(\ref{deg1}) and formula
(\ref{main}) with $z=1/2$ we find
\ba\label{Hurwitz-sum} \sum\limits_{j = 0}^{\left\lfloor {\qE } \right\rfloor }
{g_j \frac{1}{{\sqrt {\qE - j} }}} = 2\left[ {\zeta ({\textstyle{1
\over 2}},\qE  - \left\lfloor {\qE } \right\rfloor ) - \zeta
({\textstyle{1 \over 2}},\qE  + 1) - \frac{1}{{2\sqrt {\qE } }}}
\right],
\ea
where the $r.h.s.$ of this equation defines the following family of
functions that appear regularly in the thermodynamic expressions
for this system:
\ba\label{def-h} {\cal H}_z (q)\equiv \zeta (z,\left\{ q \right\})
- \zeta (z,q + 1) - \frac{1}{2}q^{ - z} ,\ea
and where $\{q\}=q-\left\lfloor q \right\rfloor$ is the fractional
part of $q$.

In what follows, we will denote energies in units
of the electron mass, $\varepsilon = E/m$, and the magnetic force
$eB$ in terms of $m^2$:
\ba\label{dimensionless-B}  b \equiv \frac{2eB}{m^2}. \ea
The variable $b$ measures the magnetic field strength in units of
the natural strength $B_0=\frac{1}{2}m^2c^4/e\hbar c \approx
2.2\times 10^{13}\,\text{Gauss}$ \cite{Bzero}.

With this notation and the use of Eq.~(\ref{Hurwitz-sum}), the density of
states for energy $E$ has a simple form:
\ba\label{gE1} g(E) = V\frac{m^2}{2\pi^2}b^{1/2}~ \varepsilon~\longhplus
\left( {\frac{\varepsilon^2  - 1 }{b}} \right). \ea
The total number of states up to a given energy $E$ is the integral
over the density of states:
\ba\label{GE0} G(E) = \int_m^E
{g(E')\,dE'}. \ea
Using Eq.~(\ref{deriv-zeta}) of Appendix A, it is easy to show that
$\longhplus(q) = 2\longhminus'(q)$ (where
the prime denotes differentiation with respect to $q$), so that we
can do this integral in a closed form, obtaining:
\ba\label{GE} G(E) =  V\frac{m^3}{2\pi^2}b^{3/2} ~~\longhminus \left(
{\frac{\varepsilon^2 - 1 }{b}} \right). \label{GE1} \ea
Clearly, $G(\Ef)/V$ is the electron density in the system, where
$\Ef$ denotes the Fermi energy. We will consistently use
dimensionless variables in the expressions:
\ba\label{dimensionless-E}
\ef \equiv \frac{\Ef}{m},\quad \pf \equiv
\sqrt{\ef^2-1}.
\ea
While $\ef$ is the Fermi energy, $\pf$ is the Fermi momentum of
the $j=0$ Landau level, in units of $m$.
The expression $\pf^2/b$ gives a measure of the occupation of
the Landau levels in the ground state: its integer
part, $\lfloor \pf^2/b \rfloor$, equals the last Landau level that
is occupied, and its fractional part $\{\pf^2/b \}$, or more
precisely $1-\{\pf^2/b \}$, gives a measure of the ``distance'' to
the next empty one.

\section{Limiting cases and regimes}

In the non-relativistic case, the energy scale associated with the
magnetic field is the spacing between Landau levels, $e B/m$,
which is uniform and equal to $\omega_B$, the cyclotron frequency.
For typical values of magnetic field strengths found in pulsars, say
$B=10^{12}$ Gauss, one finds $\omega_B\approx 10$ keV.

As is well known, a fermion gas becomes degenerate when the
occupation number falls abruptly for states with energy above a
given value $\Ef$, the Fermi energy. This happens if the
temperature is much smaller than the kinetic range $\Tf \equiv\Ef
- m$ ({\it i.e.} the energy range between the lowest orbital and
the Fermi energy). For electrons in neutron stars, typical values
might be $T < 1\unit{MeV}{}$ and $\Tf \simeq 10^2\unit{MeV}{}$, so
that the gas is not only degenerate but also highly relativistic.
In the relativistic regime however, the level spacing is not
uniform, but decreases with energy, so that close to the Fermi
surface it is $e B/\Ef$.

While the Fermi energy is invariably larger than the temperature
in a degenerate gas,  the number of occupied Landau levels may or
may not be large. This number is given by the integer part of
$(\Ef^2-m^2)/(2eB)$, in view of Eq.~(\ref{def-qE}) and its
preceding paragraph, and so it is large if $2eB/\Ef\ll \Tf$. This
is indeed the case for typical magnetic fields in neutron stars
($B<10^{14}$ G) where $eB/\Ef < 10$ keV.

We can still classify the gas in a uniform magnetic field as cold
of warm, depending on whether the temperature is smaller or larger
than the energy spacing between Landau levels close to the Fermi
surface, the latter being $eB/\Ef$.
Consequently, in a cold degenerate gas, the following condition
is satisfied:
\ba\label{cold-regime} T \ll \frac{eB}{\Ef}\quad{\rm and}\quad
T\ll \Tf. \ea
If, in addition, the magnetic field is small so that there is a
large number of occupied Landau levels, the above hierarchy
becomes:
\ba\label{cold-and-small} T \ll \frac{eB}{\Ef}\ll \Tf. \ea
On the other hand, in the warm regime, the temperature is
comparable or larger than the Landau spacing:
\ba\label{warm-regime} \frac{eB}{\Ef}< T \ll \Tf. \ea
These three cases are the regimes of interest here.

%insert figure 3 below *****************

%insert figure 3 above*********

%insert figure 4 below****************

%insert the figure in the space above

\section{The Grand Potential and Hurwitz Functions}
\label{grand-potential}

The general expression for the grand potential for an ideal Fermi
gas in a uniform background magnetic field $B$ is given by
\ba\label{Omega1} \Omega (T,B,V,\mu ) =  - T\,\sum\limits_\alpha
{\ln \left( {1 + e^{ - (E _\alpha   - \mu )/T} } \right)},
\ea
where $\mu$ is the chemical potential and the sum is over all
1-particle orbitals $\alpha$. The $B$ dependence is implicit in
the orbital energies $E_\alpha$ and the density of states.
Thermodynamic quantities can be obtained directly from $\Omega$.
In particular, $\cal M$, the magnetization per unit volume is
obtained as
\ba\label{Magnetization1} {\cal M} =  - \frac{1}{V}\frac{\partial
\Omega }{{\partial B}} . \ea
In terms of the density of states $g(E)$ the grand potential (per
unit volume) is
\ba\label{Omega3} \omega \equiv \frac{\Omega}{V}  =  -
T\,\int_m^\infty \frac{g(E)}{V}~ \ln \left( {1 + e^{ - (E - \mu
)/T} } \right)dE. \ea
This expression, together with the explicit form (\ref{gE1})
for $g(E)$, can be used to study the thermodynamics at
arbitrary temperature $T$ and magnetic field $B$.

{\bf The $T=0$ Limit}

As we show in Appendix B, for a system of non-interacting fermions
any finite temperature quantity can be obtained from its
zero-temperature counterpart.
Therefore, in the
rest of this section we will just concentrate on the grand
potential density at zero temperature,
$\omega(T=0)\equiv \omega_0$,
keeping in mind the standard definition for the Fermi energy $\Ef
\equiv\mu(T=0)$. The $T=0$ limit is simply formulated from
\begin{equation}
\lim _{T \to 0} ~ -T\ln \left( {1 + e^{ - (E - \mu )/T} } \right)=
\cases{
   E-\Ef  & for $ E < \Ef$ \cr
   0   & for $E > \Ef$,\cr}
\end{equation}
so that
\ba\label{OmegaZero1a} \omega _0 (\Ef,B) = \int_m^{\Ef}
\frac{g(E)}{V}\,E~dE - \Ef\,\frac{N}{V}. \ea
The integral in Eq.~(\ref{OmegaZero1a}), which is the ground state
energy density ($\equiv u_0$), can be done using the explicit
representation for the density of states given in Eq.~(\ref{gE1}).
After standard manipulation and integration by parts we get:
\ba u_0\equiv\int_m^{\Ef} \frac{g(E)}{V}~E\,dE &=&
\frac{m^4}{4\pi^2}\Bigg[ 2\, b^{3/2}\,\ef\,
\longhminus\left(\frac{\pf^2}{b}\right)\nn\\  &&\phantom{xxxx}-~ b^{5/2}
\int_0^{\pf^2/b}{\frac{{\longhminus (q)}}{{\sqrt {1 + bq}
\,}}\,dq}  \Bigg]  .\label{totalE} \ea
Notice that the first term above is exactly $\Ef\,N/V$, according
to Eq.\ (\ref{GE}), so it cancels with the second term in
Eq.~(\ref{OmegaZero1a}), leaving a compact form for the grand
potential density $\omega_0\equiv\Omega_0/V$:
\ba\label{OmegaZero2}\omega _0 (\ef,b) =  - \frac{m^4}{4\pi^2}
b^{5/2} \int_0^{\pf^2/b} {\frac{{\longhminus (q)}}{{\sqrt {1 + bq}
\,}}\,dq}. \ea
where
\ba\label{longhminusonehalf}
\longhminus(q)=\zeta ({-\textstyle{1 \over 2}},\left\{ q
\right\}) - \zeta ({-\textstyle{1 \over 2}},q + 1) -
\frac{1}{2}\sqrt{q}.
\ea
Formally, this is the main result of this work.

Expression (\ref{OmegaZero2}) allows us to recover the well known
non-relativistic result \cite{landau-spI} in a straightforward fashion:
in the non-relativistic limit $\pf\ll 1$, so that the square root
in the integrand of (\ref{OmegaZero2}) can simply be replaced by
unity. Using then the relation
$\longhminus(q)=\textstyle{2 \over 3}\longhminusthreehalf'(q)$ we find
\ba\label{OmegaZero2nr}\omega_0^{(\text{n.r.})} (\ef,b) =  - \frac{m^4}{6\pi^2}
b^{5/2} \longhminusthreehalf (\pf^2/b),
\ea
where
\ba\label{longhminusthreehalf} \longhminusthreehalf (q)\equiv
\zeta (-\textstyle{3 \over 2},\left\{ q \right\}) - \zeta (-\textstyle{3 \over 2},q
+ 1) - \frac{1}{2}q^{3/2}.
\ea
The result (\ref{OmegaZero2nr}) is exactly the
known non-relativistic result, after identifying
$\pf^2=2\mu_{n.r.}/m$, where $\mu_{n.r.}$ is the standard
(non-relativistic) chemical potential. In particular, notice that
$\longhminusthreehalf(q)$ splits naturally into a monotonic term
and an oscillatory term, $\longhminusthreehalf(q) =
\longhminusthreehalf^{(\text{mon})} (q) +
\longhminusthreehalf^{(\text{osc})} (q)$, as:
\ba
\longhminusthreehalf^{(\text{mon})} (q) &=& - \zeta (-{\textstyle{3 \over
2}},q + 1) - \frac{1}{2}q^{3/2} ,\\
\longhminusthreehalf^{(\text{osc})} (q)
&=&\zeta (-{\textstyle{3 \over 2}},\left\{ q \right\}).
\ea
The oscillating character of
$\longhminusthreehalf^{(\text{osc})}(q)$ arises from the
fractional part $\left\{q\right\}$ in the definition above. This
is the term that accounts for the de Haas -- van Alphen
oscillations in the non-relativistic case, and leads to Landau's
result at finite temperature, as it is shown in section
\ref{ground-state-magnetization}.

Likewise, the relativistic grand potential density
(\ref{OmegaZero2}) separates into
two terms, $\omega _0  = \omega _0^{(\text{mon})}  + \omega
_0^{(\text{osc})}$, where
\ba \omega _0^{(\text{mon})}(\ef,b) &=& \frac{m^4}{4\pi^2}b^{5/2}
\int_0^{\pf^2/b} {\frac{\zeta (-{\textstyle{1 \over 2}},q + 1) +
{\textstyle{1 \over 2}}\sqrt{q}}{{\sqrt {1 + bq} \,}}\,dq},
\label{omega0-mon-1}\\ \omega _0^{(\text{osc})}(\ef, b) &=&
-\frac{m^4}{4\pi^2} b^{5/2} \int_0^{\pf^2/b} {\frac{\zeta
(-{\textstyle{1 \over 2}},\left\{ q \right\})}{{\sqrt {1 + bq}
\,}}\,dq}. \label{omega0-osc-1} \ea
Here we have kept the notation consistent, however one must be
aware that $\omega_0 ^{(\text{osc})}$, the integral of an
oscillating function, is not purely oscillatory, as we will see.
The integral expressions above are exact and finite for any values
of the magnetic field and Fermi momentum.

Although the integrals in Eqs.~(\ref{omega0-mon-1}) and
(\ref{omega0-osc-1}) cannot be solved in a closed form, one can
nevertheless use the analytic properties of the Hurwitz zeta
function to obtain an expansion of $\omega_{0}$ for  small $b$,
or similarly for large $\pf^2/b$ ({\it i.e.} large
number of occupied Landau levels). Let us
first consider $\omega_{0}^{\rm (mon)}$. A na\"\i ve binomial
expansion of the square root in the integrand will not work
because of the diverging behavior of the integral as
$\pf^2/b\to\infty$. A procedure to obtain the correct expansion
for small $b$ consists in first extracting out a sufficient number
of leading terms in the asymptotic expansion of the zeta function
for large $q$, such that the remainder is integrable in the limit
$b\to 0$ [three terms will do for $\zeta(-1/2,q)$], and then
integrating the subtracted terms explicitly. Based on
Eq.~(\ref{asymptotic-expansion-1}) of Appendix A we write
\ba\label{zeta-sub3-1} \zeta(z,q) =   \frac{1}{{z - 1}}q^{ - z +
1}  + \frac{1}{2}q^{ - z}  + \frac{z}{{12}}q^{ - z - 1} +
\tilde\zeta_3 (z,q), \ea
implicitly defining the subtracted zeta function
$\tilde{\zeta}_{3}(z,q)$. Now, the integrals corresponding to the
exhibited terms in (\ref{zeta-sub3-1}) along with the term
$\sqrt{q}/2$ in (\ref{omega0-mon-1}) can be done exactly for
finite $b$, leaving only an integral containing the subtracted
function $\tilde \zeta _3 (-1/2,q)$:
\ba\label{OmegaZero3} \omega _0^{(\text{mon})}(\ef,b) &=&  -
\frac{m^4}{4\pi^2} \Bigg\{ \frac{1}{2} (1-b+\frac{1}{6}b^2)\ln
\left(\frac{{\ef  + \sqrt {b + \pf^2} }}{{1 + \sqrt b }}\right)
\nn\\ && \phantom{- \frac{m^4}{4\pi^2}} - \frac{1}{2} \left[
\left(b + \pf^2\right)^{1/2}\ef-b^{1/2}\right] \nn\\ &&\phantom{-
\frac{m^4}{4\pi^2}} +\frac{1}{3} \left[ \left(b +
\pf^2\right)^{3/2}\ef-b^{3/2}\right] \nn\\ &&\phantom{-
\frac{m^4}{4\pi^2}} + \frac{1}{2} b \left[ \cosh^{-1}(\ef) -
\pf\,\ef\right] \nn\\ &&\phantom{- \frac{m^4}{4\pi^2}} - b^{5/2}
\int_0^{\pf^2/b} {\frac{{\tilde \zeta _3 ( - {\textstyle{1 \over
2}},q + 1)}}{{\sqrt {1 + bq} }}} \,dq \Bigg\}.  \ea
The expression above is still exact, but now the integral
containing $\tilde \zeta _3 (z,q)$ can be expanded in powers up to
order $b$ without entering into trouble, because $\tilde \zeta _3
(-1/2,q)$ vanishes as $q^{-5/2}$ for $q\to\infty$ (see
Eq.~(\ref{asymptotic-expansion-3}) of Appendix A). It suffices to
take the upper limit to infinity and expand the denominator:
\ba\label{residual-int} \int_0^{\pf^2/b} {\frac{{\tilde \zeta _3 (
- {\textstyle{1 \over 2}},q + 1)}}{{\sqrt {1 + bq} }}} \,dq
&\buildrel {b\to 0}\over \longrightarrow& -\left(
\frac{2}{3}\zeta\left(\textstyle{-{3\over 2}}\right) +
\frac{1}{60}\right)
\\ &&+ b\left( -\frac{2}{15}\zeta\left(-\textstyle{5\over 2}\right)
+\frac{1}{1260}\right)~+~O(b^{3/2})\dots \nn \ea
This integral thus loses the dependence on the chemical potential
and becomes a spurious contribution that cancels in $\omega_0$
when we put all pieces together (see below). A word of caution is
due: to go to higher orders in $b$ one cannot continue this
expansion in the same fashion. Instead, one should follow the same
procedure from the beginning, but extracting more explicit terms
from the asymptotic expansion (\ref{zeta-sub3-1}), integrating
them directly, and then expanding the $n$-residual integral
containing $\tilde \zeta _n (-1/2,q)$ up to the last convergent
term, that goes as $b^{2n-5}$.

Now, let us turn to the oscillatory contribution $\omega
_0^{(\text{osc})}$, which can be written in the following form
(see Appendix C):
\ba\label{omega0-osc-2} \omega _0^{(\text{osc})} &=&
-\frac{m^4}{4\pi^2} b^2 \Bigg\{ \int_0^{\left\{\pf^2/b \right\}}
\frac{\zeta (-{\textstyle{1 \over 2}},q)}{\sqrt{q + 1/b +
\left\lfloor\pf^2/b \right\rfloor}} dq
\\
&\quad& + \int_0^1 \zeta (-{\textstyle{1 \over 2}},q) \left[ \zeta
\left( {\textstyle{1 \over 2}},q + {{1/ b}} \right)
 - \zeta \left(
 {\textstyle{1 \over 2}},q + {{1/ b}}
 + \left\lfloor \pf^2/b \right\rfloor \right) \right]
\, dq \Bigg\},\nn \ea
whose expansion for small $b$ is straightforward (however
tedious):
\ba\label{omega0-osc-3}\omega _0^{(\text{osc})}&\buildrel {b\to
0}\over \longrightarrow& - \frac{m^4}{4\pi^2}\Bigg\{
\frac{2}{3}b^{5/2}\left[\frac{1}{\ef}
\zeta\left(-{\textstyle{3\over 2}}, \left\{{\pf^2/
b}\right\}\right)- \zeta\left(-{\textstyle{3\over
2}}\right)\right] \\ && + \frac{2}{15}b^{7/2}\left[\frac{1}{\ef^3}
\zeta\left(-{\textstyle{5\over 2}}, \left\{{\pf^2/
b}\right\}\right)-\zeta\left(-{\textstyle{5\over
2}}\right)\right]\Bigg\}~+~O(b^{9/2})\dots
 \nn\ea
Finally we can put all the terms together and find the expansion
of the grand potential density $\omega _0$ for small magnetic
field (see Appendix D):
\ba\label{omega0-small-b-1} \omega _0 = &&-
\frac{m^4}{4\pi^2}\Bigg\{ \frac{1}{2}\cosh^{-1}(\ef)
+\frac{1}{3}\ef~\pf^3 -\frac{1}{2}\ef~\pf \nn \\ &&+~
\frac{~b^2}{12}\cosh^{-1}(\ef) +\frac{2}{3}\frac{b^{5/2}}{\ef}
\zeta\left(-{\textstyle{3\over 2}},\left\{
\pf^2/b\right\}\right)\nn \\ &&+ \frac{2}{15}\frac{b^{7/2}}{\ef^3}
\zeta\left(-{\textstyle{5\over 2}},\left\{\pf^2/b\right\}\right)\
+\  O(b^4)\dots\Bigg\}. \ea
We note that expansion (\ref{omega0-small-b-1}) is in agreement,
apart from an overall sign, with the small field result for the
effective action obtained in Ref.~\cite{cangemi} (Eq.~(74)),
which in our notation reads
\ba\label{Seff-small-b}
\left. {\Delta S^{d = 3} } \right|_{T = 0;\mu  \ne 0}  &=&
\frac{{b\,m^4 }}{{8\pi ^2 }}\Big\{ {\ef\pf - \ln (\ef + \pf)}
\Big\}\hfill
\nn\\
&&\!\!\!\!\!\!\!\!\!\!\!\!\!\!\!\!\!\!\!\!\!\!\!\!\!\!\!
\!\!\!\!\!\!\!\!\!\!\!\!\!\!\!\!\!\!\!\!\!\!\!\!\!\!\!+
\frac{{b^2 m^4 }}{{2\pi ^2 }}\sum\limits_{l = 0}^\infty
{\frac{{(b/\ef^2 )^{l + 1/2} }}{{(2l + 1)(2l + 3)}}\Big\{ {\zeta
( - l -{\textstyle{3\over 2}},\left\{\pf^2/b\right\}) - \zeta ( - l
-{\textstyle{3\over 2}},\pf^2/b)} \Big\}}
\ea
Although the non-analytic oscillatory terms are immediate to
compare, it is quite nontrivial to obtain the analytic terms from
(\ref{Seff-small-b}), since each term of the type
$b^{l+1/2}\zeta(-l-{\textstyle{3\over 2}},\pf^2/b)$ in the sum in
(\ref{Seff-small-b}) contributes to all orders in $b$ as $b\to 0$,
in view of the asymptotic expansion
(\ref{asymptotic-expansion-2}) for the Hurwitz zeta function.
Using the first three terms of this asymptotic expansion, one can
perform each of the resulting infinite sums in a closed way to
indeed obtain the terms shown in (\ref{omega0-small-b-1}), with a
vanishing coefficient for the term linear in $b$.
In retrospect, we realize that the result
(\ref{Seff-small-b}) can be formally obtained directly from the
closed expression (\ref{OmegaZero2}) by doing a binomial expansion
of the square root and then integrating term by term. However, the
latter expansion is valid over the whole integration range only if
$\pf^2<1$, {\em i.e.} basically in the non-relativistic limit.
From a numerical point of view it is clearly advantageous to use
expansion (\ref{omega0-small-b-1}) instead of
(\ref{Seff-small-b}), since then one can compute to any desired
precision at small $b$ by keeping only a finite number of terms in
the expansion.

All thermodynamic quantities at $T=0$ can be obtained from
$\omega_0$ as derivatives. One should notice that, since
$\zeta(z,0)=\zeta(z,1)$ for $z<0$, the function $\zeta(z,\{ q\})$
is continuous in $q$ for all $z<0$ and, in view of
Eq.\ (\ref{deriv-zeta}), has continuous derivative for all $z<-1$.
For example, it is straightforward to derive the expansion for the
density at $T=0$ using the thermodynamic identity
\begin{equation}
n_0=-\frac{1}{m}\frac{\partial \omega_0}{\partial \ef}.
\end{equation}
We thus find the the density at $T=0$, consistent with Eq.~(\ref{GE}), and
the corresponding expansion for small $b$:
\ba\label{exp-ne} n_0 (\ef, b)  &=& \frac{m^3}{2\pi^2}~ b^{3/2}~
\longhminus \left( \frac{\pf^2}{b} \right)\nn \\ &\to &
\frac{m^3}{2\pi^2}\Bigg\{ \frac{2}{3}\pf^3 + b^{3/2} ~\zeta\left(
- {\textstyle{1 \over 2}},\left\{\pf^2/b\right\} \right)+
\frac{b^2}{24~ \pf} ~+~ O(b^4)...\Bigg\}
 \ea
In the $b= 0$ limit we recover the free electron gas result,
$n_0 = m^3\pf^3/3\pi^2$. As seen in Eq.~(\ref{exp-ne}), the
leading correction for finite $b$ and $T=0$ is a term that
oscillates with $\pf^2/b$, corresponding to the de Haas-van Alphen
effect for the magnetization in metals.

Fig.\ 1 shows $n_0$ vs.\ $\ef$ for two values of the field.
The step-like behavior is due to the filling of consecutive
Landau levels as $\ef$ increases.
The density of states in a given Landau level [see Eq.\ (\ref{deg1})]
goes as $b\, dp_z \sim b\, dE/p_z$, for $p_z$ starting from zero. The
higher density at the bottom of each level causes the step in $n_0$.
On the other hand, as $b$ decreases the steps gradually disappear
until the smooth $b=0$ limit is reached. Alternatively, Fig.\ 2
shows $n_0$ as a function of the field $b$, for fixed Fermi
energy. Here $n_0$ oscillates as $b$ changes. Imagine we start
from a large value of $b$. As $b$ decreases, the Landau levels
move down in energy, crossing one by one the threshold defined
by $\ef$. Since the levels are denser at the bottom, $n_0$ first
grows as each level becomes accessible, but then decreases again,
because of the $b$ factor in the density of states. The
oscillation amplitude becomes smaller and smaller as the Landau levels get
closer and closer, until we reach the $b=0$ limit.

We note that the expansion in Eq.~(\ref{exp-ne}) is actually a
large $\pf^2/b$ expansion, appropriate when a large number of
Landau levels are occupied at $T=0$. So it will remain valid even
for large values of the magnetic field, say $b\sim 100$, provided
the ratio $\pf^2/b$ stays much larger than one. This will be the
case for electron densities in typical neutron stars, for which
$\pf^2\sim 10^4$.

We can also obtain the expansion for $u_0$, the energy density at
$T=0$
---whose exact integral form is given in Eq.~(\ref{totalE})---,
using the expansions in Eqs.\ (\ref{omega0-small-b-1}) and
(\ref{exp-ne}) and the thermodynamic relation
$u_0 = \omega_0 + m\,\ef\, n_0$:
\ba u_0 &=& \frac{m^4}{4\pi^2} \Bigg\{ \ef\,\pf^3 +
\frac{1}{2}\ef\,\pf-\frac{1}{2}\cosh^{-1}(\ef) \nn \\ &&
+2b^{3/2}~\ef~\zeta\left(-{\textstyle{1\over
2}},\left\{\pf^2/b\right\}\right)+ \frac{~b^2}{12}
\frac{\ef}{\pf^2} - \frac{~b^2}{12}\cosh^{-1}(\ef) \nn \\ &&
-\frac{2}{3}\frac{b^{5/2}}{\ef}\zeta\left(-{\textstyle{3\over
2}},\left\{\pf ^2/b\right\}\right)-
\frac{2}{15}\frac{b^{7/2}}{\ef^3}\zeta\left(-{\textstyle{5\over
2}},\left\{\pf^2/b\right\}\right)\nn \\ &&+~O(b^4)\dots \Bigg\}.
\ea

Unlike Eq.~(\ref{exp-ne}) for the particle density, the expansions
for the energy density given above and for the grand potential
given in Eq.\ (\ref{omega0-small-b-1}) are valid for small
$b$ only. More useful in the case of neutron star conditions is an
expansion for large $\pf^2/b$, that is, many Landau levels
occupied, but regardless of $b$. In that case, for instance,
the small $b$ expansion of the denominator in
Eq.~(\ref{residual-int}) is not valid. The correct expansion
of $\omega_0$ for large $\pf^2/b$ is the following, where
some $\pf$-independent integrals are left to be done numerically:
\ba \omega _0^{(\text{mon})} (\ef,b) &=&  - \frac{m^4}{4\pi^2}
\Bigg\{ \frac{1}{2} (1-b+\frac{1}{6}b^2)\ln \left(\frac{{\ef  +
\sqrt {b + \pf^2} }}{{1 + \sqrt b }}\right) \nn\\ && \phantom{x}
- \frac{1}{2} \left[ \left(b +
\pf^2\right)^{1/2}\ef-b^{1/2}\right] \nn\\ &&\phantom{x}
+\frac{1}{3} \left[ \left(b +
\pf^2\right)^{3/2}\ef-b^{3/2}\right] \nn\\ &&\phantom{x}
+ \frac{1}{2} b \left[ \cosh^{-1}(\ef) -
\pf\,\ef\right] \nn\\ &&\phantom{x} - b^{5/2}
\int_0^{\infty} {\frac{{\tilde \zeta _3 ( - {\textstyle{1 \over
2}},q + 1)}}{{\sqrt {1 + bq} }}} \,dq +
\frac{1}{3840}\frac{b^4}{\ef^4}+\dots\Bigg\},
\ea
\ba\omega _0^{(\text{osc})}&=& - \frac{m^4}{4\pi^2}\Bigg\{
\frac{2}{3}\frac{b^{5/2}}{\ef}
\zeta\left(-{\textstyle{3\over 2}}, \left\{{\pf^2/
b}\right\}\right)+ \frac{2}{15}\frac{b^{7/2}}{\ef^3}
\zeta\left(-{\textstyle{5\over 2}}, \left\{{\pf^2/
b}\right\}\right)+\dots\nn\\
&&\phantom{xxxxx}+\, b^2 \int_0^1 \zeta (-{\textstyle{1 \over 2}},q) \zeta
\left( {\textstyle{1 \over 2}},q + {{1/ b}} \right)\, dq\,\,
\Bigg\}.
\ea
{\bf The finite temperature case}

As shown in Appendix B, the grand potential at finite temperature
can be obtained from its expression at $T=0$, $\omega_0(\ef)$, as:
\ba\label{omega-T}
\omega(T,\mu)
=\int_{-\frac{\mu-1}{T}}^\infty \omega_0(\mu+T x)\frac{e^x}{(e^x+1)^2}
dx,
\ea
where the temperature $T$ and the chemical potential $\mu$ are
given in units of the electron mass.  If $\omega_0(\mu+T x)$ is a
slowly varying function of $x$ over a range $\Delta x \sim 1$,
then the integral in Eq.~(\ref{omega-T}) can be easily expanded in
powers of $T$:
\ba
\omega(T,\mu)= \omega_0(\mu) + \frac{\pi^2}{6}T^2
\omega_0''(\mu)+\frac{7 \pi^4}{360} T^4 \omega_0^{(4)}(\mu)+\dots.
\ea
In our case of interest, this approximation is valid for the
non-oscillatory terms $\omega_0^{(\text{mon})}$, but not for
the oscillatory terms, which vary considerably over the range
$\Delta x \sim 1$ (unless the unlikely
condition $\mu T / b\ll 1$ is met).

The treatment of the oscillatory terms, which are of the form
$\zeta (z,\{(\mu^2-1)/b\})$, for $z$ a negative semi-integer,
can be done as follows. Consider placing the leading
oscillatory term,
\ba\label{lead-osc}
\omega_0^{(\text{osc})} \approx -\frac{m^4}{4\pi^2}\Bigg\{
\frac{2}{3}\frac{b^{5/2}}{\ef}
\zeta\left(-{\textstyle{3\over 2}}, \left\{{\pf^2/b}\right\}
\right)+\dots\Bigg\},
\ea
into the integral of Eq.\ (\ref{omega-T}). Accordingly, we must
evaluate this function at $\ef'= \mu + T x$ and
${\pf'}^2\equiv {\ef'}^2-1 = \pf^2+2\mu T x + T^2 x^2$
(where $T$ is the temperature in units of the electron mass)
and weight it with the {\sl hump function} $h(x)=e^x/(e^x+1)^2$.
In the degenerate regime, $T\ll \mu$, so the term ${\cal O}(T^2)$
in the expression for ${\pf'}^2$ can be safely neglected.
Since the prefactor of the zeta function in (\ref{lead-osc})
is a slowly varying function of $x$, we can approximate
\ba\label{lead-osc2}
\omega^{(\text{osc})}(T,\mu,b) \simeq
-\frac{m^4}{6\pi^2}
\frac{b^{5/2}}{\mu}~\int_{-\infty}^\infty  {\zeta\left(-
{\textstyle{3 \over 2}},\left\{\pf^2/b  +2x~\mu T
/b\right\}\right)~ h(x)~ dx }. \ea
The number of oscillations that fall under the hump will clearly
be proportional to the factor $2\mu T/b$. We expect the amplitude
of the oscillatory magnetization to be more or less constant for
$\mu T/b < 1$ and rapidly decreasing for $\mu T/b >1$. To obtain
an explicit expression we use Hurwitz's Fourier expansion of
$\zeta(z,q)$ shown in Eq.\ (\ref{fouzeta})
(valid for $z<0$ and $0\le q\le 1$) and integrate term by term
using the formula:
\ba\label{I-zeta}
\int_{ - \infty }^\infty e^{i ax}\,h(x)dx = \frac{{\pi
a}}{{\sinh (\pi a)}}.\ea
Defining the integral in Eq.~(\ref{lead-osc2}) in
a generic form as:
\ba\label{I-Fourier}
I_z (\alpha ,\beta ) = \int_{ - \infty }^\infty {\zeta
\left(z,\{ \alpha  + \beta x\} \right) ~h(x)~ dx}, \ea
 we find it to be
\ba I_z (\alpha ,\beta ) = (2\pi )^{z+1} \Gamma (1 - z)\beta
\sum\limits_{n = 1}^\infty {\frac{{n^z \sin (2\pi n\alpha + z\pi
/2)}} {{\sinh (2\pi ^2 \beta n)}}}. \ea
Now, using this result for $z=-3/2$, the oscillatory part of
the grand potential becomes:
\ba\label{lead-osc3}
\omega^{(\text{osc})}(T,\mu,b) \simeq
+\frac{m^4}{4\pi^2}\Bigg\{
b^{3/2}\frac{T}{\sqrt{2}}\,\sum\limits_{n = 1}^\infty
{\frac{\cos (2\pi n\,\pf^2/b - \pi /4)}{n^{3/2}\,
\sinh (4\pi^2 n\,\mu\, T /b)}} + \dots \Bigg\}.
\ea

\section{The Magnetization}
\label{ground-state-magnetization}

{\bf The $T=0$ limit}

The magnetization (per unit volume) of the electron gas at
$T=0$ is given by
\ba\label{little-m-def-0} {\cal M}_0 =  - {\frac{{\partial \omega
_0(\ef,B) }}{{\partial B}}} &=&  - \frac{2e}{m^2}\times
{\frac{{\partial \omega _0(\ef,b) }}{{\partial b}}}\\ &=&
\frac{e\, m^2}{2\pi^2}~\frac{\partial }{{\partial b}}\left[
b^{5/2} \int_0^{\pf^2/b} {\frac{{\longhminus (q)}}{{\sqrt {1 + bq}
\,}}\,dq} \right].\nn \ea
Taking the derivative we find the formal expression for the
magnetization:
\ba\label{little-m-expr-1} {\cal{M}}_0(\ef , b)
&=&\frac{e\,m^2}{2\pi^2}\Bigg\{ b^{3/2} \int_0^{\pf^2/b}
{\longhminus(q)\left[ {\frac{2}{{(1 + bq)^{1/2} }} + \frac{1}{{2(1
+ bq)^{3/2} }}} \right]} \,dq \nn\\ && -~ b^{1/2}~
\frac{\pf^2}{\ef}~\longhminus\left(\pf^2/b\right)\Bigg\}. \ea
Just as $\omega_0$ was separated into two terms according to the
monotonic and the oscillating parts of $\longhminus$, so can we
separate the magnetization as ${\cal M}_0(\ef,b)={\cal
M}_0^{(\text{mon})}(\ef,b)+{\cal M}_0^{(\text{osc})}(\ef,b)$. The
small $b$ expansion for ${\cal M}_0$ can be obtained from
Eq.~(\ref{little-m-expr-1}), or directly taking the derivative of
the expansion for $\omega_0$ given in
Eq.~(\ref{omega0-small-b-1}):
\ba\label{magnetization} {\cal M}_0(\ef,b) &=&
\frac{e\,m^2}{2\pi^2}\Bigg\{
-b^{1/2}~\frac{\pf^2}{\ef}~\zeta ( - {\textstyle{1 \over 2}},\{
\pf^2/b\} + \frac{1}{6}~b~\cosh^{-1}(\ef )  )
\nn\\
&+& \frac{b^{3/2}}{3}\frac{{(4\ef^2 + 1)\,
}}{\ef^3}\zeta ( - {\textstyle{3 \over 2}},\{ \pf^2/b\} ) +
\frac{b^{5/2}}{{15}} \frac{{(4\ef^2 + 3)\, }}{\ef^5}\zeta ( -
{\textstyle{5 \over 2}},\{ \pf^2/b\} )
\nn\\
&+& O(b^3 )\dots\Bigg\}. \ea
The magnetization oscillates as a function of $b$ (see Fig.\ 3),
just like the well known de Haas-van Alphen effect of
non-relativistic electrons in metals. In this expression, the
oscillation appears in terms of Hurwitz
functions of the fractional part of $\pf^2/b$. Notice that, for
$b\lsim 1$, the oscillatory part has an amplitude considerably
larger than the monotonic part, and even larger the larger $\ef$
is (see envelope curves in Fig.\ 3).
As such, it could be possible to have spontaneous magnetization
for sufficiently dense systems at low temperature. A discussion about
the thermodynamic stability of such state was given in Ref.\
\cite{Oconnell}. However, at temperatures above some threshold,
the oscillation amplitude dies out, and with it the possibility
of spontaneous magnetization \cite{Schmid, blanford}.

\label{magnetization-finite-temperature}

{\bf The finite temperature case}

We will be concerned in particular with the physically relevant
case where $\pf^2/b\gg 1$, {\it i.e.} many Landau levels occupied.
In this case, the dominant contribution at $T=0$ is the leading
oscillatory term followed by the leading monotonic term,
{\it i.e.} the second and first terms in
Eq.~(\ref{magnetization}), respectively.

At finite temperature, the monotonic term becomes:
\ba
{\cal M}^{(\text{mon})}(T,\mu,b) &\simeq &
\frac{e\,m^2}{2\pi^2}\frac{b}{6}\Bigg\{\cosh^{-1}(\mu)
\nn\\
&&- \frac{\pi^2}{6}\frac{\mu}{\pf^3}T^2 - \frac{7\pi^4}{60}
\frac{\mu(\mu^2+{3\over 2})}{\pf^7}T^4 +\dots\Bigg\},
\ea
where $\pf^2 \equiv \mu^2 -1$. This contribution is small compared
to the oscillatory part for temperatures smaller than the Landau
level splittings, {\it i.e.} $T< b/\mu$.
However, for higher $T$ the amplitude of the oscillations gets
smoothed out and it is the monotonic part shown above what dominates
the magnetization. This behavior is shown in Fig.\ 4.

On the other hand, the leading oscillatory term at finite $T$ is
obtained following the same procedure as in the previous section,
or directly taking the derivative of the grand potential
with respect to $b$:
\ba\label{magnetization-T-osc} {\cal M}^{(\text{osc})}(T,\mu,b) =
-\frac{e\,m^2}{\pi}\frac{\pf^2\, T}{\sqrt{2b}}\,\sum\limits_{n =
1}^\infty {\frac{\sin (2\pi n\,\pf^2/b - \pi /4)}{\sqrt {n}\,
\sinh (4\pi^2 n\,\mu\, T /b)}}+\dots. \ea
This expression coincides with the results in Ref.\ \cite{blanford}.
We must also remark the strong similitude with Landau's result for the
non-relativis\-tic case \cite{landau-spI}, which is expressed in our
notation as:
\ba\label{magnetization-non-rel} {\cal
M}^{(\text{osc})}_{n.r.}(T,\mu,b) = -\frac{e\,m^2}{\pi}\frac{\pf^2\,
T}{\sqrt{2b}}\,\sum\limits_{n = 1}^\infty {\frac{\sin (2\pi
n\,\pf^2/b - \pi /4)}{\sqrt {n}\, \sinh (4\pi ^2 n\, T/b)}}. \ea
This is precisely the non-relativistic limit of our leading
oscillating term, shown in Eq.~(\ref{magnetization-T-osc}):
basically, the relativistic chemical potential reduces to
$\mu\equiv 1+\mu_{n.r.} \to 1$, in units of electron mass,
while $\pf \equiv \sqrt{\mu^2-1} \to \sqrt{2\mu_{n.r.}}$
denotes a non-trivial quantity in all regimes.

A final important point is the possibility of ferromagnetic
behavior, that is, a magnetization that is sustained with the
magnetic field generated by the same system. For a given geometry,
the field and magnetization satisfy a relation of the form
$M=\gamma B$, where in the case of {\it e.g.} a sphere, $\gamma =
3/8\pi$. A self-consistent solution is obtained from this relation
and the thermodynamic relation $M = M(T,\mu,B)$. This problem has
been studied in the past, and our conclusions
agree with those results. For the self-consistent solution
to exist, the oscillations are necessary, because the monotonic
part of $M$ vs. $b$ is simply too small to reach a solution
with $M=\gamma B$. Moreover, for the oscillations not to
be thermally damped out, the temperature must be below some
threshold value \cite{Schmid}. Finally, there is the question of
thermodynamic stability of the self-consistent solution, which
is at most metastable according to O'Connell and
Roussel \cite{Oconnell}.

\bigskip
\noindent{\Large \bf Conclusions}

We have developed a closed analytical approach to solve the
thermodynamics of a free gas of electrons immersed in an
uniform magnetic field of arbitrary magnitude. The method is
completely relativistic and particularly useful in the case of a
degenerate gas, a likely situation to be met in the interior of
white dwarfs, neutron stars, and magnetars. A central role is played by
one of the least known of the special functions of mathematical
physics, the Hurwitz zeta function. Its appearance comes about
due to a sum over energy levels of the form $(\alpha+\beta n)^z$
(Landau levels for the system treated in this paper), labeled by a
non-negative integer $n$. As such, the method should be applicable
to study the thermodynamics of other systems.
The grand potential (and therefore all thermodynamic quantities) can
be expressed as a one-dimensional definite integral, which can be
explicitly evaluated in several interesting regimes due to the various
analytic properties of the Hurwitz zeta function.
Hence, our work provides a unified derivation of several of the
results found scattered in the literature. We reproduce the
de Haas-van Alphen behavior of the magnetization in the relativistic
gas and the dilution of it at high temperatures.

We have not included quantum electrodynamics corrections in this
treatment. Work to extend our results in this direction, in a fully
relativistic fashion, is in progress.

\bigskip
\noindent{\Large \bf Acknowledgments}
\medskip
\ret This work was supported by CONICYT under Grant
Fondecyt PLC-8000017.

\newpage
\noindent{\Large \bf Appendix A: The Hurwitz Zeta Function}
\label{hurwitz-zeta-function}

We collect here some of the properties of the Hurwitz zeta
function and present a (presumably original) derivation of its
asymptotic behavior for large $q$. For a detailed account consult,
for instance, references \cite{whittaker-watson} or \cite{atlas}.

The Hurwitz zeta function $\zeta(z,q)$ is defined as the analytic
extension to the whole complex $z$ plane of the series
\ba \zeta(z,q) & = &\sum_{n=0}^{\infty} \frac{1}{(n+q)^{z}} \ea
valid for $\text{Re } z > 1$ and $q \neq 0, -1, -2, \dots$. The
resulting function is analytic everywhere except at $z=1$, where
it has a simple pole with unit residue.

For $q\ne 0$ one has
\ba \zeta(z,q+1)= \zeta(z,q) - \frac{1}{q^{z}}, \label{zetastar}
\ea
which, iterated $N$ times, leads to
\ba\label{main} \sum_{n=0}^{N} \frac{1}{(n+q)^{z}} & = &
\zeta(z,q) - \zeta(z,q+N+1). \ea
This finite sum is what appears in the grand potential as a sum
over Landau levels, and its expression in terms of Hurwitz Zeta
functions is what allows us to expand the thermodynamic quantities
in different limiting scenarios.

The derivative of $\zeta(z,q)$ with respect to $q$ is again a
Hurwitz zeta function:
\ba\label{deriv-zeta} \frac{\partial }{{\partial q}}\zeta (z,q) =
- z\,\zeta (z + 1,q). \ea
Hermite's integral representation \cite{whittaker-watson}
\ba\label{hermite} \zeta(z,q) & = & \frac{1}{z-1}q^{-z+1} +
\frac{1}{2} q^{-z} + 2\,q^{-z+1} \int_{0}^{\infty} \frac{\sin( z
\tan^{-1} t) dt}{(1+t^2)^{z/2} \left( e^{2 \pi t q} -1 \right) },
\ea
valid for all $z\ne 1$ and $q>0$, can be used to study the large
$q$ behavior of $\zeta(z,q)$. For large $q$, the leading
contribution to the integral in Eq.~(\ref{hermite}) comes from the
region of small $t$. To isolate this contribution we split the
integration range into $[0,a]$ and $[a,\infty)$, where $a$ is a
fixed number less than one (for instance, $a=1/2$). In order to
approximate the first integral we shall use the remarkable series
expansion:
\ba (1 + t^2 )^{ - z/2} \sin (z\tan^{-1} (t)) = \sum\limits_{k =
0}^\infty  {( - 1)^k } \frac{{(z)_{2k + 1} }} {{(2k + 1)!}}t^{^{2k
+ 1} },\label{series} \ea
which converges uniformly for $\abs{t}<1$. In (\ref{series})
$(z)_{n}$ is the Pochhammer symbol, or shifted factorial, defined
by
\ba (z)_n= \frac{\Gamma(z+n)}{\Gamma(z)}= z(z+1)\cdots (z+n-1).
\ea
Thus, for a given $z$ and $0<t\le a < 1$ we have the uniform
approximation
\ba\label{series-2} \left| {\frac{{\sin (z\tan ^{ - 1} t)}}{{t\,(1
+ t^2 )^{z/2} }} - \sum\limits_{k = 0}^N {( - 1)^k \frac{{(z)_{2k
+ 1} }}{{(2k + 1)!}}t^{2k} } } \right| < \varepsilon, \ea
where $\varepsilon$ can be an arbitrarily small positive number
and $N=N(\varepsilon, z)$ is a sufficiently big number. So we can
write
\ba\label{app-int-1} \int_0^a {\frac{{\sin (z\tan ^{ - 1} t)}}{{(1
+ t^2 )^{z/2} (e^{2\pi qt}  - 1)}}dt = \sum\limits_{k = 0}^N {( -
1)^k \frac{{(z)_{2k + 1} }}{{(2k + 1)!}}\int_0^a {\frac{{t^{2k +
1} }}{{e^{2\pi qt}  - 1}}\,} } } dt + R(\varepsilon ), \ea
with
\ba\label{bound-R} \left| {R(\varepsilon )} \right| < \int_0^a
{\frac{\varepsilon t}{{e^{2\pi qt}  - 1}}\,} dt < \varepsilon
\int_0^\infty {\frac{t}{{e^{2\pi qt}  - 1}}\,} dt =
\frac{\varepsilon }{{24q^2 }}. \ea
Also, up to a correction that vanishes exponentially as
$q\to\infty$, we can approximate
\ba\label{app-int-2} \int_0^a {\frac{{t^{2k + 1} }}{{e^{2\pi qt} -
1}}\,} dt \simeq \int_0^\infty  {\frac{{t^{2k + 1} }}{{e^{2\pi qt}
- 1}}\,} dt = ( - 1)^k \frac{{B_{2k + 2} }}{{4(k + 1)q^{2k + 2}
}}, \ea
where we have used formula 3.411.2. of reference \cite{gr}. The
$B_n$ in (\ref{app-int-2}) are the Bernoulli numbers of even
index, $B_0=1,\;B_2=1/6,\;B_4=-1/30$, etc. Since the integral over
$[a,\infty)$ in (\ref{hermite}) also gives a exponentially small
contribution for large $q$, we finally have the result
\ba\label{app-zeta-1} \zeta (z,q) &=& \frac{1}{{z - 1}}q^{ - z +
1}  + \frac{1}{2}q^{ - z}  + q^{ - z - 1} \sum\limits_{k = 0}^N {(
- 1)^k \frac{{(z)_{2k + 1} }{B_{2k + 2} }}{{(2k + 2)!{q^{2k} }}}}
\nn\\ &&+ O(\varepsilon ) + O(e^{ - \alpha q} ). \ea
Letting $\varepsilon\to 0$ (and consequently $N\to\infty$) we
obtain the following expansion for $\zeta(z,q)$:
\ba \zeta (z,q) &=& \frac{1} {{z - 1}}q^{1 - z}  + \frac{1} {2}q^{
- z}  + \sum\limits_{k = 0}^\infty  {\frac{{B_{2k + 2}}} {(2k +
2)!}} (z)_{2k+1} \frac{1} {{q^{z + 2k + 1} }}
\label{asymptotic-expansion-1}\\ & = & \frac{1} {{\Gamma
(z)}}\sum\limits_{k = 0}^\infty {( - 1)^k \frac{{B_k }} {{k!}}}
\frac{{\Gamma (k + z - 1)}} {{q^{k + z - 1} }},
\label{asymptotic-expansion-2} \ea
where (\ref{asymptotic-expansion-2}) follows because the only
non-vanishing Bernoulli number of odd index is $B_1=-1/2$.

The cases $z=-1/2$ and $z=1/2$ are of special relevance for the
work of this paper and we give the corresponding asymptotic
expansions explicitly:
\ba \label{asymptotic-expansion-3} \zeta(-{\textstyle{1 \over
2}},q) &=&  - \frac{2}{3}q^{3/2}  + \frac{1}{2}q^{1/2}  -
\frac{1}{{24}}q^{ - 1/2}  + O(q^{ - 5/2} ),\\
\label{asymptotic-expansion-4} \zeta ({\textstyle{1 \over 2}},q)
&=&  - 2q^{1/2}  + \frac{1}{2}q^{ - 1/2}  + \frac{1}{{24}}q^{ -
3/2}  + O(q^{ - 7/2} ). \ea
For general $z$, the expansion (\ref{asymptotic-expansion-1}) is
only asymptotic. It is easy to check that its radius of
convergence, as a series in the variable $1/q$, is zero. However,
as an aside we should mention that for $z=-m$, where $m$ is a
non-negative integer, the series (\ref{asymptotic-expansion-1})
terminates, in view of $(-m)_n=0$ for $n>m$. Hence the series
becomes a finite polynomial in $q$, actually a Bernoulli
polynomial, up to a multiplicative constant:
\ba \zeta(-m,q)=-\frac{1}{m+1}B_{m+1}(q),\quad m=0,1,2,\cdots. \ea
The Hurwitz zeta function for $z<0$ admits the following Fourier
expansion in the range $0<q<1$:
\begin{equation}
\zeta(z,q) = \frac{2 \Gamma(1-z)}{(2 \pi)^{1-z}}  \times
\left( \sin\left( \frac{ \pi z}{2} \right) \sum_{n=1}^{\infty}
\frac{\cos( 2 \pi q n)}{n^{1-z} }  +
\cos\left( \frac{ \pi z}{2} \right) \sum_{n=1}^{\infty}
\frac{\sin( 2 \pi q n)}{n^{1-z} }  \right).
\label{fouzeta}
\end{equation}

\bigskip
\noindent{\Large \bf Appendix B: Finite Temperature from T = 0}
\label{finite-temperature}

In this appendix we show that the full finite temperature grand
potential shown in Eq.~(\ref{Omega3}) can be obtained from its
zero temperature limit as
\ba\label{omega-t0-a} \Omega (T,\mu ) = \int_{ - \frac{{\mu  - m}}
{T}}^\infty  {\Omega _0 (\mu  + Tx)\,h(x)dx}. \ea
where $h(x)$ is the Fermi-Dirac hump,
\ba\label{FDhump-a} h(x)
= \frac{{e^x }}{{(e^x  + 1)^2 }} = \frac{1}{{4\cosh ^2 (x/2)}}.
\ea
Actually, result (\ref{omega-t0-a}) is a special case of the
fact that any physical quantity of the form
\ba\label{Phys-Quantity-finite-T} Q(T,\mu )  =  \int_m^\infty
{q(E)\frac{1}{{e^{(E - \mu )/T}  + 1}}dE} \ea
can be computed in terms of its zero temperature limit
$Q_0(\mu)$ as
\ba\label{Phys-Quantity-finite-T-2} Q(T,\mu ) = \int_{ -
\frac{{\mu  - m}} {T}}^\infty  {Q _0 (\mu  + Tx)\,h(x)dx}. \ea
To prove (\ref{Phys-Quantity-finite-T-2}) we express the function
$q(E)$ in Eq.~(\ref{Phys-Quantity-finite-T}) as the derivative of
a function $F(E)$ defined as:
\ba F(E) \equiv \int_m ^E q(E') dE' ,\ea
and integrate by parts. After the change of variable $E=\mu+Tx$ we
obtain
\ba\label{Omega3pp2} Q(T,\mu )  = \int_{ - \frac{{\mu  -
m}}{T}}^\infty  {F(\mu  + Tx)\frac{{e^x }}{{(e^x + 1)^2 }}dx}. \ea
But from (\ref{Phys-Quantity-finite-T}) it is seen that the
function $F(\mu)$ is precisely the value of $Q(T,\mu)$ in the
limit $T\to 0$:
\ba Q_0(\mu)=\lim _{T \to 0} Q(T,\mu ) = \int_m^{\mu} {q(E)dE
\equiv F(\mu)}. \ea
This establishes (\ref{Phys-Quantity-finite-T-2}). Now, to prove
Eq.~(\ref{omega-t0-a}) for the grand potential, we merely notice
that the standard expression for it shown in Eq.~(\ref{Omega3})
can be turned into the form (\ref{Phys-Quantity-finite-T}) by
expressing the density of states $g(E)$ as the derivative of the
function $G(E)$ --which is the total number of single particle
states up to energy $E$:
\ba\label{GE0-a} G(E) = \int_m^E {g(E')\,dE'},
\ea
and then integrating by parts to obtain
\ba\label{Omega3pp1} \Omega(T,\mu ) =  - \int_m^\infty
{G(E)\frac{1}{{e^{(E - \mu )/T} + 1}}dE}. \ea
This expression for the grand potential is precisely of the form
shown in Eq.~(\ref{Phys-Quantity-finite-T}), with $q(E)=-G(E)$.

\medskip

In the degenerate regime, $T\ll\mu-m$, the lower limit of
integration in (\ref{omega-t0-a}) can be replaced by $-\infty$
with negligible error. Additionally, if the function $\Omega _0
(\mu + Tx)$ varies slowly under the hump (this may not be the case
for the oscillatory terms; see section
\ref{magnetization-finite-temperature}), then it can be expanded
in a Taylor series around $\mu$ and the resulting terms integrated
one by one. Using now the results
\ba \int_{-\infty}^\infty  h(x)~dx &=& 1,\quad \int_{-\infty
}^\infty x^2~ h(x)~dx = \frac{\pi ^2 }{3}, \quad \int_{-\infty
}^\infty x^4~ h(x)~dx = \frac{7\pi ^4 }{15},\nn \\
&& \dots  \int_{-\infty
}^\infty x^{2n}~ h(x)~dx = \pi^{2n} \left|(2^{2n}-2)\, B_{2n}\right|  ,\ea
we obtain the small temperature expansion,
\ba\label{Omega-t-a} \Omega(T,\mu) = \Omega_0(\mu) +
\frac{~\pi^2}{6}~T^2\ \Omega_0''(\mu)+ \frac{~7\pi^4}{360}~T^4\
\Omega_0^{(4)}(\mu)\dots \ea

\bigskip
\noindent{\Large \bf Appendix C: Reduction of the Oscillatory Contribution}
\label{oscillatory-contribution}

Let $p(q)$ be a function in the unit interval $[0,1]$ periodically
extended over the real axis, and $f(q)$ an arbitrary function.
Then, the integral of the product of these functions over an
arbitrary interval can be separated into a sum of integrals over
integer intervals, plus a residual integral, as follows:
\[
\int_0^Q {p(q)f(q)\,dq = \int_0^1 {p(q)\sum\limits_{k =
0}^{\left\lfloor Q \right\rfloor  - 1} {f(q + k)\,dq +
\int_0^{\left\{ Q \right\}} {p(q)f(q + \left\lfloor Q
\right\rfloor } )\,dq} } }.
\]
We now specialize this result for $p(q) = \zeta(z,\{q\})$ as the
periodic function, $f(q)=(1+bq)^{-s}$ and $Q=\pf^2/b$, and use
Eq.~(\ref{main}) to express the sum in terms of Hurwitz functions:
\ba \sum\limits_{k = 0}^{\left\lfloor {\pf^2/b} \right\rfloor  -
1} {\frac{1}{{\left[ {1 + b(q + k)} \right]^s }}}  &=&
\frac{1}{{b^s }}\sum\limits_{k = 0}^{\left\lfloor {\pf^2/b}
\right\rfloor  - 1} {\frac{1}{{\left[ {(q + 1/b) + k}  \right]^s
}}} \nn\\ & = & \frac{1}{{b^s }}\left[ {\zeta \left(s,q +
1/b\right) - \zeta \left(s,q + 1/b + \left\lfloor\pf^2/b
\right\rfloor \right)} \right]. \nn \ea
We thus find the integral expression
\ba \label{oscillating}\int_0^{\pf^2/b} {\frac{{\zeta \left(z,\{
q\} \right)}}{{(1 + bq)^s }}\,dq} &=& \frac{1}{{b^s }}\Bigg\{
\int_0^{\{ \pf^2/b\}}{\frac{{\zeta (z,q)}}{{\left( {q + 1/b +
\left\lfloor \pf^2/ b \right\rfloor } \right)^s }}\,dq}
\\ &&\!\!\!\!\!\!\!\!\!\!\!\!+
\int_0^1 \zeta (z,q)\left[ {\zeta \left( s,q + 1/b\right) - \zeta
\left( s,q + 1/b + \left\lfloor\pf^2/b \right\rfloor \right)}
\right]\,dq \Bigg\}.\nn \ea
Using this generic result one derives the expression given in
Eq.~(\ref{omega0-osc-2}) for the oscillatory piece of the grand
potential.

\bigskip
\noindent{\Large \bf Appendix D: Small $b$ Expansions}
\label{small-b-expansion}

In this appendix we show that
\ba\label{small-b-exp-1} - \int_0^{\pf^2/b} {\frac{{\tilde \zeta
_3 ( - {\textstyle{1 \over 2}},q + 1)}}{{\sqrt {1 + bq} }}} \,dq
~~=~ \frac{2}{3}\zeta ( - {\textstyle{3 \over 2}}) +
\frac{1}{{60}} \!\!\!\!&+&\!\!\! b
 \left[ {\frac{2}{{15}}\zeta ( - {\textstyle{5 \over 2}}) -
\frac{1}{{1260}}} \right]\nn \\&&\quad +~ O(b^{3/2} )\dots \ea
and
\ba\label{small-b-exp-2} \int_0^{\pf^2/b} {\frac{{\zeta ( -
{\textstyle{1 \over 2}},\{ q\} )}}{{\sqrt {1 + bq} }}}
 dq &=& - \frac{2}{3}\zeta \left( -
{\textstyle{3 \over 2}}\right) - \frac{2}{{15}}\zeta \left( -
{\textstyle{5 \over 2}}\right)\,b - \frac{2}{{35}}\zeta \left( -
{\textstyle{7 \over 2}}\right)\,b^2 \nn\\ &&+~
\frac{2}{3}\frac{1}{\ef}\zeta \left( - {\textstyle{3 \over 2}},\{
\pf^2/b\} \right) + \frac{2}{{15}}\frac{b}{\ef^3}\zeta \left( -
{\textstyle{5 \over 2}},\{ \pf^2/b\} \right) \nn\\ &&+~
\frac{2}{{35}}\frac{{b^2 }}{\ef^5}\zeta \left( - {\textstyle{7
\over 2}},\{ \pf^2/b\} \right) + O(b^3 )\dots \ea
The first expansion is needed in order to find the small $b$
behavior of the non-oscillatory piece of the grand potential,
Eq.~(\ref{OmegaZero3}), and the second expansion is required for
the oscillatory piece, which is exhibited in
Eq.~(\ref{omega0-osc-2}) already making use of the formula
(\ref{oscillating}).

To prove Eq.~(\ref{small-b-exp-1}), we first define the integral
\ba R(b)\equiv  - \int_0^{\pf^2/b} {\frac{{\tilde \zeta _3 ( -
{\textstyle{1 \over 2}},q + 1)}}{{\sqrt {1 + bq~} }}} \,dq ,\nn
\ea
and then try to expand it in powers of $b$. The first two terms of
the expansion are easily found, using the results:
\ba \left. {R(b)} \right|_{b = 0}  &=& \frac{2}{3}\zeta ( -
{\textstyle{3 \over 2}}) + \frac{1}{{60}}, \nn\\ \left.
{\frac{\partial }{{\partial b}}R(b)} \right|_{b = 0}  &=&
\frac{2}{{15}}\zeta ( - {\textstyle{5 \over 2}}) -
\frac{1}{{1260}}, \nn \ea
which follow from the facts that ${\tilde \zeta _3 ( -
{\textstyle{1 \over 2}},q + 1)}$ decreases like $q^{-5/2}$ when
$q\to\infty$, and that the antiderivative of $\tilde \zeta _3 ( -
{\textstyle{1 \over 2}},q)$, which is ${\textstyle{2\over
3}}\tilde \zeta _3 ( -{\textstyle{3 \over 2}},q)$, vanishes like
$q^{-3/2}$ when $q\to\infty$. Indeed, we can directly calculate at
$b=0$:
\ba \left. {R(b)} \right|_{b = 0}  =  - \int_1^\infty {\tilde
\zeta _3 ( - {\textstyle{1 \over 2}},q)} \,dq  &=&
\frac{2}{3}\tilde \zeta _3 ( - {\textstyle{3 \over 2}},1)\nn\\ &=&
\frac{2}{3}\zeta ( - {\textstyle{3 \over 2}}) + \frac{1}{{60}},
\ea
and also
\ba \left. {\frac{\partial }{{\partial b}}R(b)} \right|_{b = 0}
&=& \lim _{b \to 0} \Bigg[ \tilde \zeta _3 \left( - {\textstyle{1
\over 2}},1 + \pf^2/b\right) ~\frac{\pf^2}{\ef~b^2} \nn\\
&&\phantom{xxxxxxxxxxxxx} + \frac{1}{2}\int_0^{\pf^2/b} {\tilde
\zeta _3 ( - {\textstyle{1 \over 2}},q + 1)\frac{q}{{(1 +
bq)^{3/2} }}} \,dq \Bigg] \nn\\ &=& \frac{1}{2}\int_0^\infty
q\,\tilde \zeta _3 ( - {\textstyle{1 \over 2}},q + 1)\,dq. \nn\\
\ea
In this last expression, the $b\to 0$ limit of the first term
inside the square bracket vanishes due to the large-$q$ behavior
of $\tilde \zeta _3 ( - {\textstyle{1 \over 2}},q)$, while the
last integral can be done by parts:
\ba \frac{1}{2}\int_0^\infty  q\,\tilde \zeta _3 ( - {\textstyle{1
\over 2}},q + 1)\,dq&=& \frac{1}{2}\left( { - \frac{2}{3}}
\right)\int_0^\infty  {\tilde \zeta _3 ( - {\textstyle{3 \over
2}},q + 1)\,d} q \nn\\ &=& \frac{1}{3} \cdot \frac{2}{5}~\tilde
\zeta _3 ( - {\textstyle{5 \over 2}},1) ~=~ \frac{2}{{15}}\zeta (
- {\textstyle{5 \over 2}}) - \frac{1}{{1260}}. \nn \ea
We should point out that the next term in the expansion
(\ref{small-b-exp-1}) is not $O(b^2)$, as it would be in a regular
Taylor series, because $R''(0)$ does not exist. To get this
singular term one should go back to the original definition of
$R(b)$ and subtract the leading term in the asymptotic expansion
of $\tilde \zeta _3 ( - {\textstyle{1 \over 2}},q)$:
\ba \tilde \zeta _3 ( - {\textstyle{1 \over 2}},q) =
\frac{1}{1920\,q^{5/2} } + \tilde \zeta _4 ( - {\textstyle{1 \over
2}},q) , \ea
where $\tilde \zeta _4 ( - {\textstyle{1 \over 2}},q)$ is of order
$O(q^{-9/2})$. Then, the resulting term in the integrand,
$(1+q)^{-5/2}(1+bq) ^{-1/2}$, can be integrated explicitly, and
its small $b$ expansion reads
\ba \int_0^{\pf^2/b} {\frac{1}{{(1 + q)^{5/2} \sqrt {1 + bq} }}}
\,dq = \frac{{\rm 2}}{{\rm 3}} - \frac{2}{3}b +\frac{2}{3}
\frac{\ef}{\pf^3} (2\,\pf^2 - 1)~ b^{3/2}  + O(b^2 ) \ea
Therefore, the $O(b^{3/2})$ term in the expansion of $R(b)$ is
actually
\ba -\frac{1}{2880}~\frac{\ef}{\pf^3}~(2\,\pf^2 - 1)~b^{3/2}. \ea
With this result we complete the proof of
Eq.~(\ref{small-b-exp-1}).

We now proceed to prove Eq.~(\ref{small-b-exp-2}), which is
relevant for the small $b$ expansion of the oscillatory piece of
the grand potential. First we must realize that the integral in
Eq.~(\ref{small-b-exp-2}) is of the type studied in Appendix C,
and in particular the result shown in Eq.~(\ref{oscillating})
applies. We will then proceed to expand that result for small $b$.
To expand the first integral in Eq.~(\ref{oscillating}), we just
need to use the binomial expansion of $ \left( {q + {\textstyle{1
\over b}} + \left\lfloor \pf^2/b \right\rfloor } \right)^{ -
1/2}$, where $q \ll 1/b + \lfloor \pf^2/b \rfloor $, and then use
the integrals:
\ba \int_0^Q \zeta ( - {\textstyle{1 \over 2}},q)\,dq &=&
\frac{2}{3}\zeta ( - {\textstyle{3 \over 2}},Q) - \frac{2}{3}\zeta
( - {\textstyle{3 \over 2}}),  \\ \int_0^Q q\,\zeta ( -
{\textstyle{1 \over 2}},q)\,dq &=&  - \frac{4}{{15}}\zeta ( -
{\textstyle{5 \over 2}},Q) + \frac{4}{{15}}\zeta ( - {\textstyle{5
\over 2}}) + \frac{2}{3}Q\,\zeta ( - {\textstyle{3 \over 2}},Q).
\ea
We thus find the expansion for the first integral of
Eq.~(\ref{oscillating}):
\ba \int_0^{\{ \pf^2/b\} } {\frac{{\zeta ( - {\textstyle{1 \over
2}},q)}}{{\left( {q + 1/b + \left\lfloor \pf^2/b \right\rfloor }
\right)^{1/2} }}\,dq}
\phantom{xxxxxxxxxxxxxxxxxxxxxxxxx}\nn \ea
\ba \label{exp-incomplete} =&&\!\!\!\!\!\!\!\frac{1}{{(1 +
b\left\lfloor {\pf^2/b} \right\rfloor )^{1/2} }}\left[
{\frac{2}{3}\zeta ( - {\textstyle{3 \over 2}},\{\pf^2/b\} ) -
\frac{2}{3}\zeta ( - {\textstyle{3 \over 2}})} \right] \nn\\ &-&
\!\!\!\!\frac{b}{{(1 + b\left\lfloor {\pf^2/b} \right\rfloor
)^{3/2} }}\bigg[  - \frac{2}{{15}}\zeta ( - {\textstyle{5 \over
2}},\{ \pf^2/b\} ) + \frac{2}{{15}}\zeta ( - {\textstyle{5 \over
2}})\nn  \\
&&\qquad\qquad\qquad\qquad +~ \frac{1}{3}\{ \pf^2/b\} \,\zeta ( -
{\textstyle{3 \over 2}},\{ \pf^2/b\} ) \bigg] ~~+~~ O(b^2 )\dots
 \ea
Notice that the terms of the form $(1 + b\left\lfloor {\pf^2/b}
\right\rfloor )^{ - n/2}~ \zeta ( - {\textstyle{{n + 2} \over 2}})
$ will cancel in the full oscillatory piece,
Eq.~(\ref{omega0-osc-2}), when expansions (\ref{exp-incomplete})
and (\ref{exp-complete}) (see below) are combined. For the
remaining terms containing $(1 + b\left\lfloor {\pf^2/b}
\right\rfloor )$ in the denominator we write $1 + b\left\lfloor
{\pf^2/b} \right\rfloor  = (1 + \pf^2) - b\{ \pf^2/b\}$ and
perform a binomial expansion, which leads to a further
cancellation of all the terms of the type $\{ \pf^2/b\}^k \zeta(z,
\{ \pf^2/b\})$ with $k\ge 1$.

Now we need to expand the second integral in
Eq.~(\ref{oscillating}). In this case we must expand the $\zeta$
functions inside the squared brackets for large values of their
argument. We thus use the large-$q$ expansions of
$\zeta(\pm\textstyle{1\over 2},q)$ shown in
Eqs.~(\ref{asymptotic-expansion-3}) and
(\ref{asymptotic-expansion-4}), together with the definite
integrals \cite{Espinosa-Moll}
\ba \int_0^1 \zeta ( - {\textstyle{1 \over 2}},q)\,dq &=& 0,  \\
\int_0^1 q\,\zeta ( - {\textstyle{1 \over 2}},q)\,dq &=&
\frac{2}{3}\zeta ( - {\textstyle{3 \over 2}}),  \\ \int_0^1 q^2
\zeta ( - {\textstyle{1 \over 2}},q)\,dq &=& \frac{2}{3}\zeta ( -
{\textstyle{3 \over 2}}) - \frac{8}{{15}}\zeta ( - {\textstyle{5
\over 2}}).
\ea
We thus get the expansion:
\ba \frac{1}{{\sqrt b }} \int_0^1 \zeta ( - {\textstyle{1 \over
2}},q)\left[ {\zeta ({\textstyle{1 \over 2}},q + 1/b) - \zeta
({\textstyle{1 \over 2}},q + 1/b + \left\lfloor \pf^2/b
\right\rfloor )} \right]\,dq  \phantom{xxxxxxxxx}\nn \ea
\ba\label{exp-complete} &=& -~\frac{2}{3}\zeta ( - {\textstyle{3
\over 2}})\left[ {1 - \frac{1}{{(1 + b\left\lfloor {\pf^2/b}
\right\rfloor )^{1/2} }}} \right] - \frac{2}{{15}}\zeta ( -
{\textstyle{5 \over 2}})\left[ {1 - \frac{1}{{(1 + b\left\lfloor
{\pf^2/b} \right\rfloor )^{3/2} }}} \right]\,b \nn\\ &&+~~ O(b^2
)\dots \ea
which completes the proof of Eq.~(\ref{small-b-exp-2}) up to order
$b$. The $O(b^2)$ terms are obtained in a similar fashion.

Finally, and for completeness, we also present here the small $b$
expansion of the explicit terms in (\ref{OmegaZero3}), which reads
\ba
&&\frac{1}{2} (1-b+\frac{1}{6}b^2)\ln \left(\frac{{\ef  +
\sqrt {b + \pf^2} }}{{1 + \sqrt b }}\right)
 - \frac{1}{2} \left[ \left(b + \pf^2\right)^{1/2}\ef-b^{1/2}\right]
\nn\\
&&\quad+~\frac{1}{3} \left[ \left(b + \pf^2\right)^{3/2}\ef-b^{3/2}\right]
 + \frac{1}{2} b \left[ \cosh^{-1}(\ef) - \pf\,\ef\right]
\nn\\ &&\phantom{xxxxxxxxxxxx} =~ \frac{1}{2}\cosh ^{-1}(\ef) -
\frac{1}{2}\pf\,\ef +\frac{1}{3}\pf^3\,\ef  +
\frac{1}{{12}}b^2\cosh ^{-1} (\ef ) \nn
\\ &&\phantom{xxxxxxxxxxxxxx}
- \frac{1}{{60}}b^{5/2}  + \frac{1}{{1260}}b^{7/2} ~+~
O(b^{9/2})\dots \ea
One should notice that all the terms that are independent of the
Fermi energy ({\it i.e.} $\pf$ or $\ef$) are spurious, cancelling
between the different expansions and thus leading to result
(\ref{omega0-small-b-1}).

%\newpage
%{\Large\bf Figure Captions}

%{\parskip=12pt
%{\bf Figure 1:} {The particle density at T=0 as a function of the Fermi energy, for fixed magnetic field
%$b=0.1$ (smooth curve) and $b=0.5$ (bumpy curve).}

%{\bf Figure 2:} {The particle density at T=0 as a function of the magnetic field, for fixed
%Fermi energy $\ef=1.3$ (lower curve) and $\ef=1.5$ (upper curve).}

%{\bf Figure 3:} {The magnetization at $T=0$ which oscillates as a function of
%the magnetic field $b$, for Fermi energy $\ef = 4$. Also shown are
%the upper and lower envelopes of the curve, and the upper envelopes
%for larger Fermi energies ($\ef = 8$ and $16$).  }

%{\bf Figure 4:} {The upper envelope of the magnetization as a function of the
%temperature, for fixed chemical potential $\mu=100$, and magnetic field
%$b= 0.1,\, 1$ and $10$ (lower, medium and upper curve, respectively).}

%{\bf Figure 5:} {The upper envelope of the magnetization as a function of the
%magnetic field $b$, for two temperatures $T=10^{-5}$ and $10^{-7}$
%(lower and upper curve, respectively). The chemical potential is fixed
%at $\mu = 100$. The dashed line represents the macroscopic relation
%$M = \frac{8\pi}{3} B$, for a uniformly magnetized sphere.}

%{\bf Figure 6:} {The functions $\zeta(z,\{q\})$, for $z=-1/2, -3/2, -5/2$
%(large, medium and small amplitude, respectively).}

\newpage
\begin{figure}[h!]
%\vspace{-2 cm}
\hspace{-0.5 cm}
\mbox{ \epsfxsize=12 cm\epsffile{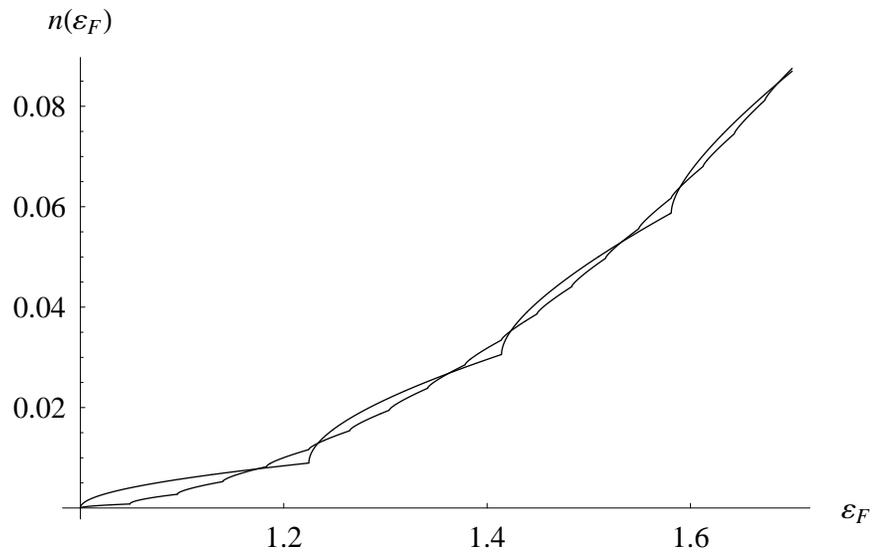}}

\vspace{2 cm}
\caption{The particle density at T=0 as a function of the Fermi energy, for fixed magnetic field
$b=0.1$ (smooth curve) and $b=0.5$ (bumpy curve).}
\end{figure}

\newpage
\begin{figure}[h!]
%\vspace{-2 cm}
\hspace{-0.5 cm}
\mbox{ \epsfxsize=12 cm\epsffile{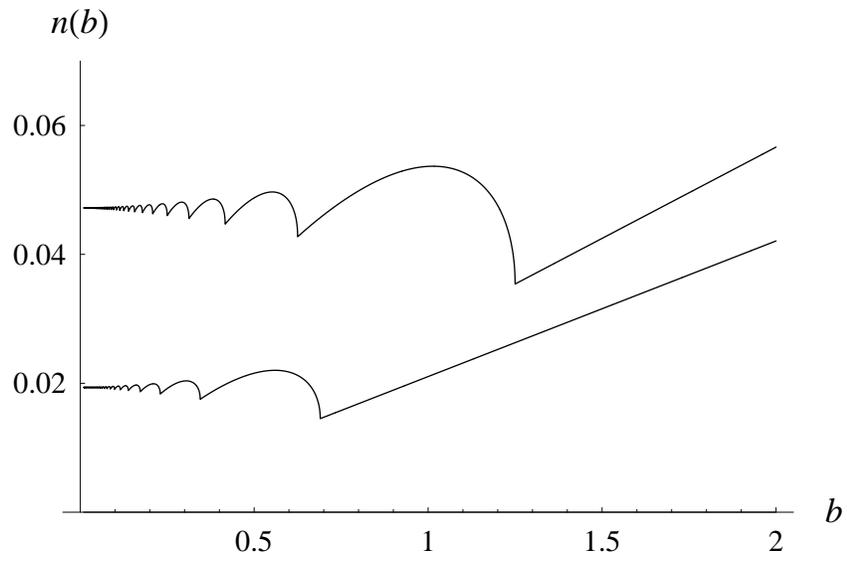}}

\vspace{2 cm}
\caption{The particle density at T=0 as a function of the magnetic field, for fixed
Fermi energy $\ef=1.3$ (lower curve) and $\ef=1.5$ (upper curve).}
\end{figure}

\newpage
\begin{figure}[h!]
%\vspace{-2 cm}
\hspace{-0.5 cm}
\mbox{ \epsfxsize=12 cm\epsffile{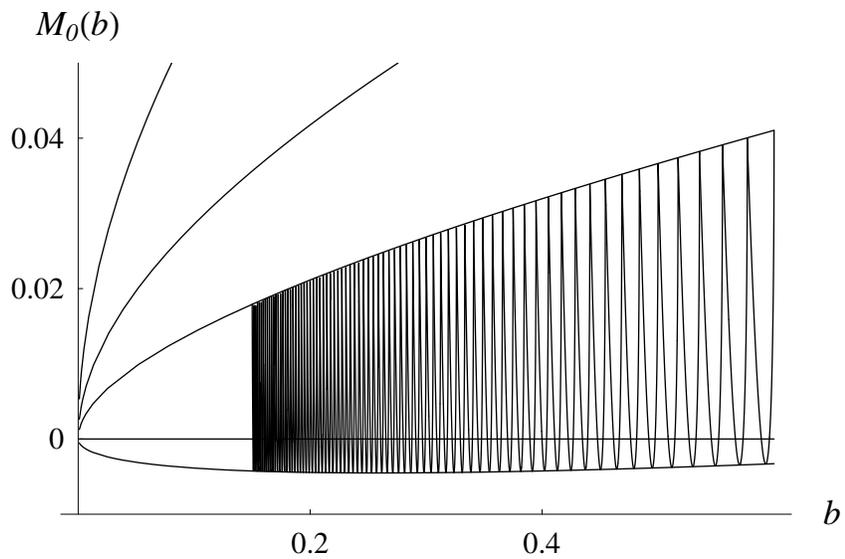}}

\vspace{2 cm}
\caption{The magnetization at $T=0$ which oscillates as a function of
the magnetic field $b$, for Fermi energy $\ef = 4$. Also shown are
the upper and lower envelopes of the curve, and the upper envelopes
for larger Fermi energies ($\ef = 8$ and $16$).  }
\end{figure}

\newpage
\begin{figure}[h!]
%\vspace{-2 cm}
\hspace{-0.5 cm}
\mbox{ \epsfxsize=12 cm\epsffile{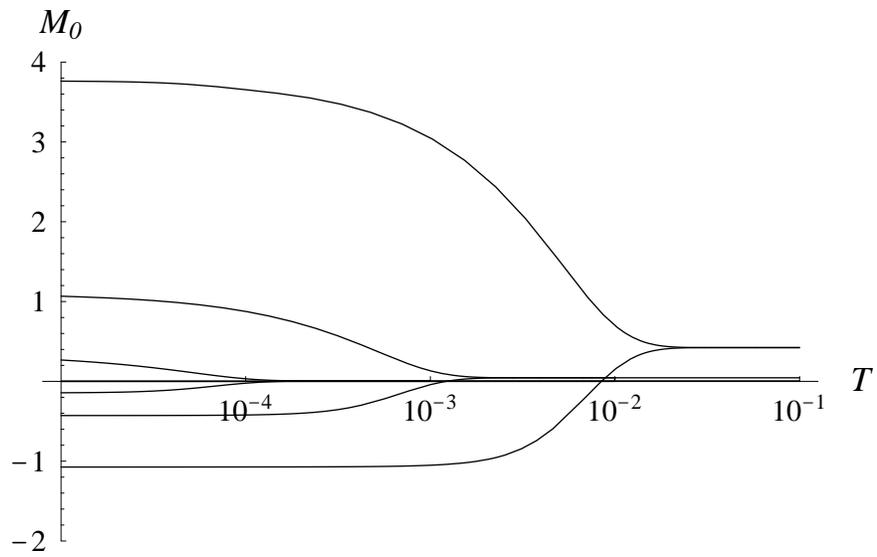}}

\vspace{2 cm}
\caption{The upper envelope of the magnetization as a function of the
temperature, for fixed chemical potential $\mu=100$, and magnetic field
$b= 0.1,\, 1$ and $10$ (lower, medium and upper curve, respectively).}
\end{figure}

\newpage
\begin{figure}[h!]
%\vspace{-2 cm}
\hspace{-0.5 cm}
\mbox{ \epsfxsize=12 cm\epsffile{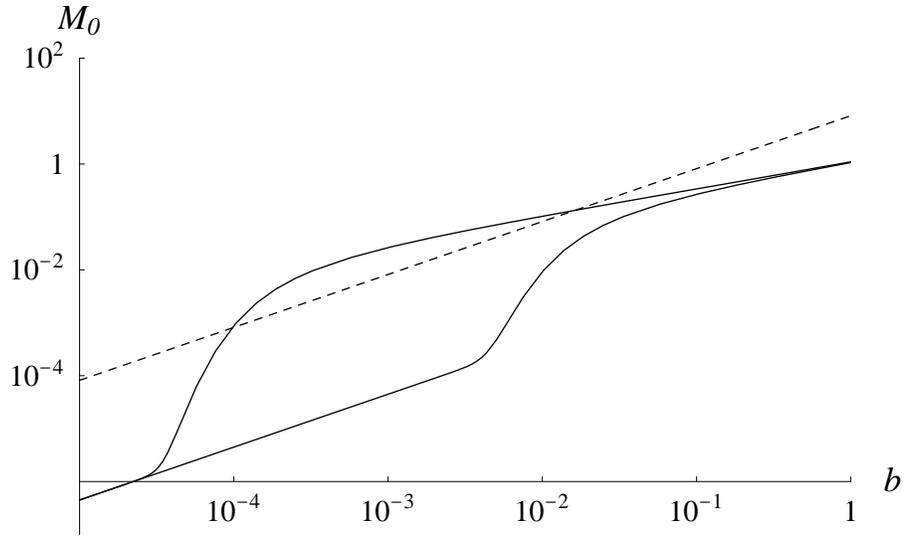}}

\vspace{2 cm}
\caption{The upper envelope of the magnetization as a function of the
magnetic field $b$, for two temperatures $T=10^{-5}$ and $10^{-7}$
(lower and upper curve, respectively). The chemical potential is fixed
at $\mu = 100$. The dashed line represents the macroscopic relation
$M = \frac{3}{8\pi} B$, for a uniformly magnetized sphere.}
\end{figure}

\newpage
\begin{figure}[h!]
%\vspace{-2 cm}
\hspace{-0.5 cm}
\mbox{ \epsfxsize=12 cm\epsffile{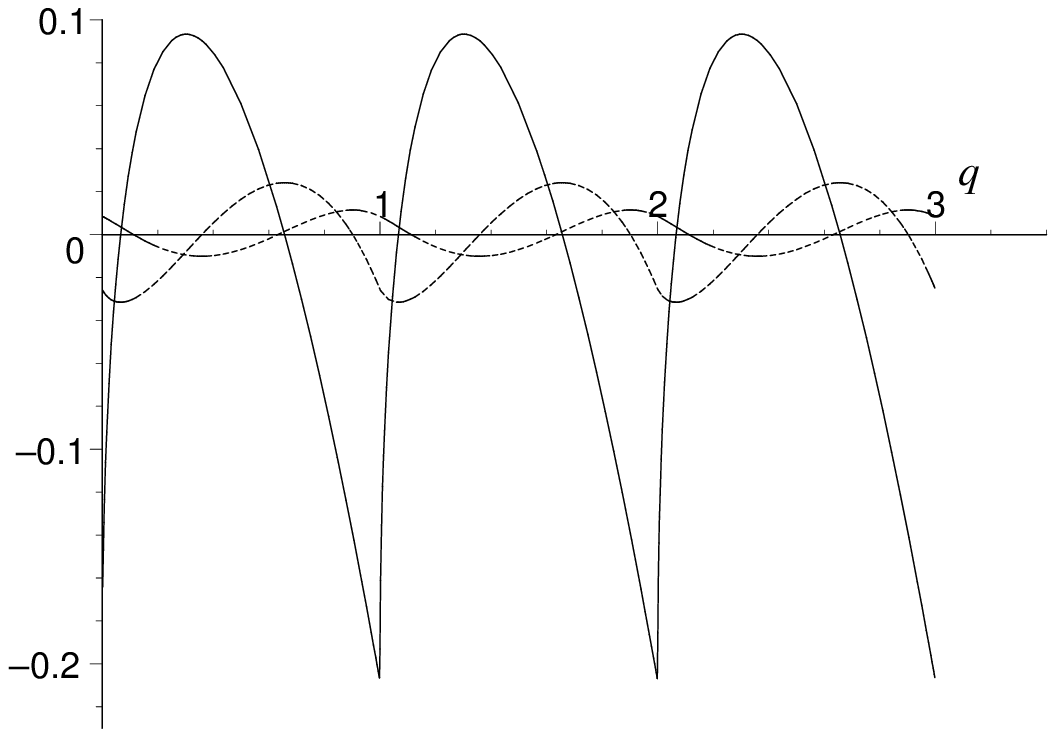}}

\vspace{2 cm}
\caption{The functions $\zeta(z,\{q\})$, for $z=-1/2, -3/2, -5/2$
(large, medium and small amplitude, respectively).}
\end{figure}


\newpage
\begin{thebibliography}{99}
\bibitem{Canuto}
V.\ Canuto and H.\ Chiu, Phys.\ Rev.\ 173, 1210 (1968); {\em ibid.}
173, 1220 (1968); {\em ibid.} 173, 1229 (1968); H.J.\ Lee, V.\
Canuto, H.\ Chiu and C.\ Chiuderi, Phys.\ Rev.\ Lett.\ 23, 390 (1969);
V.\ Canuto, H.\ Chiu and C.\ Chiuderi, Nature 225, 47 (1970).

\bibitem{Oconnell}
R.F.\ O'Connell and K.M.\ Roussel, Astron.\ \& Astroph.\
{\bf 18}, 198 (1972).

\bibitem{Schmid}
J.\ Schmid-Burgk, Astron.\ \& Astroph. {\bf 26}, 335 (1973).

\bibitem{blanford}
R.D.\ Blanford and L.\ Hernquist, J.\ Phys.\ {\bf C15}, 6233 (1982).

\bibitem{chodos}
A.~Chodos, K.~Everding and D.~Owen, Phys.~Rev.~{\bf D42}, 2881 (1990).

\bibitem{blau}
S.K.~Blau, M.~Visser and A.~Wipf, Intl.~Jour.~Mod.~Phys.~{\bf A6}
5409 (1991).

\bibitem{elmfors}
P.~Elmfors, D.~Persson and B-S.~Skagerstam, Phys.~Rev.~Lett.~{\bf 71}, 480
(1993); Astropart.~Phys.~{\bf 2}, 299 (1994).

\bibitem{zeitlin}
Vad.~Zeitlin, hep-ph/9412204 (unpublished);
D.~Persson and Vad.~Zeitlin, Phys.~Rev.~{\bf D51}, 2026 (1995);
Vad.~Zeitlin, J.~Exp.~Theor.~Phys.~{\bf 82}, 79 (1996).

\bibitem{cangemi}
D.~Cangemi and G.~Dunne, Annals~Phys.~{\bf 249}, 582 (1996).

\bibitem{landau-paper1}
L.~D.~Landau, Zeitschrift f\"{u}r Physik {\bf 64}, 629 (1930).

\bibitem{landau-paper2}
L.~D.~Landau, Proc.~Royal~Soc. of London,~{\bf A170}, 363 (1939).

\bibitem{landau-spI} See, for example, L.D.\ Landau and E.M.\ Lifshitz,
{\em Satistical Physics, Part I.}, 3rd.\ Edition, Pergamon
Press (1980).

\bibitem{lippman}
M.H.\ Johnson and B.A.\ Lippmann, Phys.\ Rev.\ 76, 828 (1949); H.\ Robl
Acta Phys.\ Austriaca 6, 105 (1952).

\bibitem{landau-qm}
L.D.\ Landau and E.M.\ Lifshitz, {\em Quantum Mechanics
(Non-relativistic Theory)}. Course of Theoretical Physics, volume
3, third edition. Pergamon Press, 1977.

\bibitem{Bzero} Notice that $B_0$ is in our case half of the scale
used by other authors \cite{Canuto, blanford}
($m^2c^3/e\hbar\approx 4.4\times 10^{13}$ G), because the natural
dimensionless variable associated to the magnetic field in this
problem is $b = B/B_0 = 2eB/m^2$, as seen in
Eqs.~(\ref{OmegaZero0}, \ref{eqforEf0}).

\bibitem{whittaker-watson}
E.\ Whittaker and  G.\ Watson, {\em A course of Modern Analysis},
Cambridge University Press, Fourth Edition reprinted, 1963.

\bibitem{atlas}
J.\ Spanier and K.B.\ Oldham, {\em An Atlas of Functions},
Hemisphere Publishing Corp., 1987.

\bibitem{gr}
I.S.\ Gradshteyn and I.M.\ Ryzhik, {\em Table of Integrals, Series
and Products}, fifth edition, ed. Alan Jeffrey. Academic Press,
1994.

\bibitem{Espinosa-Moll} O.~Espinosa and V.~Moll, {\em On some
definite integrals involving the Hurwitz Zeta function}, April 2000,
to appear in The Ramanujan Journal.

\end{thebibliography}
\end{document}